\documentclass[11pt]{imag-ms-template}
\usepackage[utf8]{inputenc}
\usepackage{graphicx}
\usepackage{caption}
\usepackage{subcaption}
\usepackage{adjustbox}
\usepackage{booktabs}
\usepackage{multirow}

\def\bfone{\mathbf 1}

\def\bfx{\mathbf x}
\def\bfy{\mathbf y}

\def\bfX{\mathbf X}

\def\bfY{\mathbf Y}

\def\bftheta{\boldsymbol \theta}
\def\bfTheta{\boldsymbol \Theta}
\def\bfalpha{\boldsymbol \alpha}
\def\bfeta{\boldsymbol \eta}

\def\bfepsilon{\boldsymbol \epsilon}

\def\bfbeta{\boldsymbol \beta}

\def\log{\text{log}}

\def\rest{\mathrm{rest}}
\def\Var{\mathrm{Var}}
\def\Cov{\mathrm{Cov}}

\title{SIMBA: Scalable Image Modeling using a Bayesian Approach, A Consistent Framework for Including Spatial Dependencies in fMRI Studies
}

\author{Yuan Zhong$^1$, Gang Chen$^2$, Paul A. Taylor$^2$ and Jian Kang$^1$\\[2mm]
$^1$University of Michigan, Ann Arbor\\
$^2$National Institute of Mental Health, NIH, Bethesda, MD, USA}

\author{Yuan Zhong$^{1}$,  Gang Chen$^{2}$,  Paul A. Taylor$^{2}$, Jian Kang$^{1\ast}$\\
{\small $^{1}$Department of Biostatistics, University of Michigan, Ann Arbor, USA}\\
{\small $^{2}$Scientific and Statistical Computing Core, National Institute of Mental Health, NIH, Bethesda, MD, USA}\\
{\small $^\ast$Correspondence:  jiankang@umich.edu}
}

\begin{document} 

\maketitle


\begin{abstract}
Bayesian spatial modeling provides a flexible framework for whole-brain fMRI analysis by explicitly incorporating spatial dependencies, overcoming the limitations of traditional massive univariate approaches that lead to information waste. In this work, we introduce SIMBA, a Scalable Image Modeling using a Bayesian Approach, for group-level fMRI analysis, which places Gaussian process (GP) priors on spatially varying functions to capture smooth and interpretable spatial association patterns across the brain volume. To address the significant computational challenges of GP inference in high-dimensional neuroimaging data, we employ a low-rank kernel approximation that enables projection into a reduced-dimensional subspace. This allows for efficient posterior computation without sacrificing spatial resolution, and we have developed efficient algorithms for this implemented in Python that achieve fully Bayesian inference either within minutes using the Gibbs sampler or within seconds using mean-field variational inference (VI). Through extensive simulation studies, we first show that SIMBA outperforms competing methods in estimation accuracy, activation detection sensitivity, and uncertainty quantification, especially in low signal-to-noise settings. We further demonstrate the scalability and interpretability of SIMBA in large-scale task-based fMRI applications, analyzing both volumetric and cortical surface data from the NARPS and ABCD studies.
\end{abstract}

\section{Introduction}
Functional magnetic resonance imaging (fMRI) has emerged as one of the most widely used neuroimaging techniques for noninvasive mapping of brain activity. In typical task-based fMRI studies, participants are scanned while performing specific cognitive or motor tasks, with the primary goal of identifying brain regions that are activated in response to the experimental conditions and often response contrasts. By measuring blood-oxygen-level-dependent (BOLD) signals, fMRI provides rich spatial and temporal information on brain function. These data are indexed by spatial locations, which can be represented either in a volumetric format, where responses are measured at voxels across the three-dimensional brain volume, or in a cortical surface format, where responses are mapped onto two-dimensional surface meshes of the cortex. In both cases, these spatial locations can be denoted using three-dimensional coordinates to describe their relative positions within the brain volume.

Traditional task-based fMRI analysis studies operate in a massive univariate analysis (MUA) framework. First, a time series regression model is independently fitted to each spatial location on single-participant data to obtain estimated coefficients associated with task effects or contrasts between different conditions. These coefficients comprise the individual-specific effect estimates or contrast maps, which are then used for subsequent or group-level analyses to map brain activations. Then, MUA is applied at the group level, where each spatial location is treated independently, and univariate hypothesis tests such as $t$-tests or general linear models (GLMs) are applied at each location. This approach is often followed by multiple testing adjustment procedures of statistic values, such as family-wise error (FWE) rate adjustment or false discovery rate (FDR) control. Although MUA is computationally efficient and widely used in popular neuroimaging software such as FSL (FMRIB Software Library; \citet{jenkinson2012fsl}), SPM (Statistical Parametric Mapping; \citet{ashburner2012spm}), and AFNI (Analysis of Functional NeuroImages; \citet{cox1996afni}), it has critical limitations. MUA results in information waste by ignoring spatial correlations and common information shared across voxels, as it assumes independence across spatial locations and lacks model flexibility. They also operate solely on statistical values, ignoring the importance of the effect estimates from modeling \citep{Chen2017-qt}. Furthermore, multiple testing adjustment methods commonly applied in MUA pipelines often cause information loss by producing binary decisions rather than providing full inference on the magnitude and uncertainty of activation effects ~\citep{marchini2004comparing, chen2019handling}.

To address these limitations, it is crucial to account for the spatial dependence inherent in fMRI data, within a consistent and appropriate statistical modeling framework. Bayesian models with spatial priors provide a principled approach to encode this dependence, allowing models to borrow strength across neighboring locations and improve both model performance and interpretability. Popular methods include conditional autoregressive (CAR) models and Gaussian Markov random fields (GMRFs), which impose local spatial smoothness through sparse precision matrices ~\citep{descombes1998spatio, woolrich2004constrained, mejia2020bayesian, zhang2016spatiotemporal, siden2017fast}. Other extensions incorporate mixture models with Dirichlet process clustering to account for latent spatial heterogeneity ~\citep{jbabdi2009multiple, zhang2016spatiotemporal}. More recently, Gaussian process (GP) priors have gained more attention in neuroimaging as a flexible method to model globally smooth spatial functions, with various applications in modeling activation regions, individual-specific effects, and spatially structured residual noise ~\citep{whiteman2024bayesian, lin2024latent}.

In general, Bayesian spatial models face significant computational challenges in whole-brain fMRI analyses, primarily due to the high dimensionality of the data, with the number of spatial locations often exceeding $10^5$. Fully Bayesian inference via Markov chain Monte Carlo (MCMC) sampling becomes computationally infeasible, especially when accounting for spatial dependence across the entire brain. To mitigate this issue, some studies have focused on region-based analyses, either restricting inference to small regions of interest (ROIs) ~\citep{hartvig2000spatial} or averaging voxel-level effects within predefined anatomical regions ~\citep{chen2019handling}, both of which sacrifice the fine spatial resolution inherent in voxel-wise fMRI data. 

To improve scalability, approximate inference methods have been widely explored. Many studies adopt variational inference (VI), which offers a computationally efficient alternative to MCMC by approximating the posterior through optimization ~\citep{zhang2016spatiotemporal, penny2005bayesian, siden2017fast}. Another useful direction is based on the integrated nested Laplace approximation (INLA). ~\citet{mejia2020bayesian} introduced spatial Bayesian GLMs that combine INLA with GMRF priors, taking advantage of sparse precision matrices to achieve substantial computational gains. This approach has been shown to deliver accurate posterior estimates and can scale to tens of thousands of voxels or surface vertices with runtimes on the order of hours. However, because GMRF priors are defined through the sparsity structure of the precision matrix, the induced spatial dependence is limited to local neighborhoods. This restriction makes them well-suited for local smoothing but less flexible for capturing global or long-range spatial correlations. 


GP priors offer a powerful alternative for modeling globally smooth and spatially varying effects. However, exact inference under GP priors involves dense covariance matrix inversions with computational complexity $\mathcal{O}(V^3)$ and memory cost $\mathcal{O}(V^2)$, where $V$ is the number of spatial locations. This makes traditional MCMC approaches infeasible for whole-brain fMRI analyses, particularly for Hamiltonian Monte Carlo (HMC) methods commonly used in software like Stan ~\citep{carpenter2017stan} and PyMC ~\citep{salvatier2016probabilistic}. ~\citet{whiteman2024bayesian} proposed a two-stage MCMC-based framework using Vecchia approximations for cortical surface fMRI data. ~\citet{lin2024latent} developed a population-weighted image-on-scalar regression method using low-rank GP kernel approximations, with a maximum likelihood estimation method for parameter estimation instead of Bayesian inference. Despite these advances, there remains a lack of a unified, fully Bayesian framework capable of performing computationally efficient inference across the whole brain volume using GP priors.
 
To address these challenges, we propose a Scalable Image Model using a Bayesian Approach (SIMBA) for modeling spatial effects in group-level fMRI analysis. This is a unified Bayesian modeling framework with GP priors that can efficiently perform whole-brain level analysis on fMRI data. The SIMBA model explicitly targets effect estimation at each spatial location across the brain volume, while simultaneously incorporating participant-specific effects and spatially correlated residuals. This framework allows for flexible and statistically consistent modeling of inter-participant variability with spatial dependencies, while also enabling posterior predictive checks as a principled method for validating model fit and assessing the adequacy of spatial inferences, such as goodness of fit, detection of systematic misfit, assumption checking, sensitivity to priors, and model comparisons~\citep{gabry2019visualization}. 

To reduce computational complexity in the implementation of this method, we adopt a low-rank approximation of the GP covariance kernel, projecting the high-dimensional spatial functions to a lower-dimensional subspace. Our approach shares a similar kernel decomposition and model reparameterization strategy with ~\citet{lin2024latent}, extending it to full Bayesian inference and allowing for more flexible kernel choices. We implement efficient posterior inference through both Gibbs sampling and variational inference, enabling whole-brain Bayesian spatial inference at voxel-level resolution within minutes or seconds. This makes our method feasible for high-dimensional fMRI data, which has previously been unattainable in standard neuroimaging practice. We demonstrate the performance of our approach through three example cases: first, with simulation studies where results can be compared with known results; and then with real fMRI volumetric data from the Neuroimaging Analysis Replication and Prediction Study (NARPS)~\citep{botvinik2020variability} in a whole brain analysis and finally with a gray matter-focused analysis of real data from the Adolescent Brain Cognitive Development (ABCD) study ~\citep{casey2018adolescent} to identify cortical activation patterns. 

The remainder of this paper is organized as follows. In Section~\ref{sec:methods}, we introduce the SIMBA framework and the corresponding prior specifications. In Section~\ref{sec:post_comp}, we demonstrate the kernel approximation, model reparameterization, and Bayesian inference algorithms, including the Gibbs sampler and variational inference. In Section~\ref{s:simulation}, we present simulation studies that evaluate model performance, followed by a real data application in Section~\ref{sec:real_data_apps} on task-based fMRI volumetric data from the NARPS project and cortical gray matter surface data from the ABCD study. We conclude with a discussion in Section~\ref{sec:discuss}.

\section{Methods}\label{sec:methods}
\subsection{Model Specification}
Suppose fMRI data are collected from $N$ individuals, each measured on $V$ spatial locations across the brain. Let $\mathcal{S} = \{s_v\}_{v=1}^V\subset\mathbb{R}^d$ denote the $d$-dimensional coordinates of the spatial locations of the volumetric voxels or surface vertices. 
For simplicity of notation, we refer to all locations as voxels indexed by $v$ throughout. Additionally, while we use a single index to label voxels, the original geometric spatial structure is maintained and used within the modeling process. 

We denote $y_i(s_v)$ as the observed image response at the spatial location $s_v$ and let $x_{ij}$ represent the $j$-th covariate for participant $i$, such as demographic or clinical information, with $i = 1, \ldots, N$ and $j = 0, \dots, J$. The proposed SIMBA model is specified as:
\begin{equation}
\label{e:IM}
\begin{split}
    y_i(s_v) &= \sum^J_{j=0}x_{ij}\alpha_j + \sum^J_{j=0}x_{ij}\beta_j(s_v) + \eta_i(s_v) + \epsilon_i(s_v), \quad \epsilon_i(s_v) \sim \mathcal{GP}(0, \sigma^2_\epsilon K).
\end{split}
\end{equation}
We include $x_{i0}=1$ for all $i=1, \dots, N$ to model the global intercept $\alpha_0$ and the spatially varying intercept $\beta_0(s_v)$ for baseline brain activity patterns. Here, $\alpha_j$ represents the global effect of the $j$th covariate shared across all locations, and $\beta_j(s_v)$ captures the spatially varying population-level effect of the same covariate at location $s_v$. Participant-specific deviations from the voxel-level population effect are represented by $\eta_i(s_v)$, and $\epsilon_i(s_v)$ denotes spatial noise, which is assumed to follow a zero-mean Gaussian process with a covariance kernel $K$ and unknown kernel variance $\sigma^2_\epsilon$. To ensure model identifiability, we impose the constraints: $\sum_{i=1}^N \eta_i(s_v) = 0$ for all $v = 1, \dots, V$, and $\sum^V_{v=1}\beta_j(s_v)=0$ for $j=0, \dots, J$. Our primary parameter of interest is the voxel-level effect 
$\alpha_j + \beta_j(s_v)$ for $j=0, \dots, J$, which combines both the global and spatial varying association between the intercept or the $j$-th covariate and brain activities. 

For each global effect $\alpha_j$, we assign a normal prior with mean zero and an unknown variance $\sigma_\alpha^2$, denoted as $\alpha_j \sim \mathcal{N}(0, \sigma_\alpha^2)$. We incorporate spatial information by assigning GP priors with zero mean and the same predefined covariance function $K(\cdot, \cdot)$ as for the residuals $\epsilon_i(s_v)$ to the latent spatial functions 
$\beta_j(\cdot) \sim \mathcal{GP}(0, \sigma^2_\beta K)$ and $\eta_i(\cdot) \sim \mathcal{GP}(0, \sigma^2_\eta K)$. For computational simplicity, we assume a shared covariance kernel $K$ across all three latent spatial functions, while allowing for different kernel variances $\sigma^2_\beta, \sigma^2_\eta$ and $\sigma^2_\epsilon$ to capture varying degrees of spatial variation across the respective model components. In practice, different kernels could be specified for the individual-level effects $\eta_i(s_v)$ if desired. The choice of kernel function determines the smoothness of spatially varying functions and spatial correlations across voxels. In this work, we use a Mat\'ern kernel, which is widely adopted in spatial statistical modeling in fMRI studies~\citep{mejia2020bayesian, siden2021spatial} due to its flexibility in controlling both correlation range and smoothness. Details on the kernel construction and approximation strategies are provided in Section~\ref{ss:kernel}. We place half-Cauchy priors on the unknown standard deviations as $\sigma_\alpha, \sigma_\beta, \sigma_\eta, \sigma_\epsilon \sim \mathcal{C}^+(A)$, where $A=100$ induces weakly informative priors. To enable efficient posterior sampling with conjugate priors, we reparameterize the half-Cauchy prior using the inverse gamma (IG) distribution as $\sigma_\alpha^2 \sim \mathrm{IG}(\frac{1}{2}, \frac{1}{a_\alpha}), \sigma_\beta^2 \sim \mathrm{IG}(\frac{1}{2}, \frac{1}{a_\beta}), \sigma_\eta^2 \sim \mathrm{IG}(\frac{1}{2}, \frac{1}{a_\eta}), \sigma_\epsilon^2 \sim \mathrm{IG}(\frac{1}{2}, \frac{1}{a_\epsilon})$, where $a_\alpha, a_\beta, a_\eta, a_\epsilon \sim \mathrm{IG}(\frac{1}{2}, \frac{1}{A^2})$. This hierarchical formulation preserves conjugacy in the posterior computations while retaining the desirable heavy-tailed shrinkage property of the half-Cauchy prior.

\subsection{Kernel approximation}
\label{ss:kernel}
A key challenge in applying GP priors to whole-brain fMRI data lies in the computational cost. Inference using GP priors requires the inversion of dense covariance matrices, which incurs $\mathcal{O}(V^3)$ in complexity. This makes standard GP inference infeasible for voxel-level analysis, where $V$ often exceeds $100{,}000$. To address this, we adopt a low-rank approximation strategy based on the Nystr\"om method~\citep{nystrom1930praktische}. This approach not only significantly reduces computational complexity but also preserves the flexibility to incorporate any parametric covariance kernel.

Let $K_V =K(\mathcal{S}, \mathcal{S}) \in \mathbb{R}^{V \times V}$ denote the Gram matrix over all spatial locations $\mathcal{S}$. We select $L \ll V$ inducing points $\mathcal{S}_L \subset \mathcal{S}$ (a process described below in Sec.~\ref{ss:repara}), and define the reduced matrices $K_L = K(\mathcal{S}_L, \mathcal{S}_L) \in \mathbb{R}^{L \times L}, \quad K_{V, L} = K(\mathcal{S}, \mathcal{S}_L)\in \mathbb{R}^{V \times L}.$
Following the Nystr\"om method approximation strategy, the full kernel can be approximated as
$K_V \approx K_{V, L} K_L^{-1} K_{V, L}^\top$, where the inducing points serve as a representative set that captures the dominant spatial structure, enabling a computationally efficient representation of the kernel.

In the next step, we represent this approximated matrix $K_V$ through a spectral decomposition for subsequent model inference. We first compute the Cholesky decomposition $K_L = R_L R_L^\top$, where $R_L \in \mathbb{R}^{L \times L}$ is a lower triangular matrix.
Plugging this into the Nystr\"om expression yields: $K_V \approx K_{V, L} R_L^{-\top} R_L^{-1} K_{V, L}^\top = \widetilde{K} \widetilde{K}^\top, $ where $\widetilde{K} = K_{V, L} R_L^{-\top} \in \mathbb{R}^{V \times L}$. We then apply singular value decomposition (SVD) on $\widetilde{K}$ to get $\widetilde{K}=\Psi DU^\top$ where  $\Psi\in \mathbb{R}^{V\times L}$ is an orthonormal matrix, i.e. $\Psi^\top\Psi=I_L$, where $I_L\in \mathbb{R}^{L\times L}$ is the $L$-dimensional identity matrix. $D\in \mathbb{R}^{L\times L}$ is a diagonal matrix containing the singular values of $\widetilde{K}$, and $U\in \mathbb{R}^{L\times L}$ is another orthonormal matrix. Then the approximated kernel can be written as $K_V \approx \Psi \Lambda\Psi^\top$ with $\Lambda = D^2=\text{diag}(\lambda_1, \dots, \lambda_L)$, which aligns with the kernel decomposition in Mercer's theorem. The obtained matrices $\Psi$ and $\Lambda$, which contain the approximated basis functions and their corresponding eigenvalues from the spectral decomposition, are then used to reparameterize the regression model formulation, as described in Section \ref{ss:repara}. For the kernel approximation of the individual-specific deviations $\eta_i(\cdot)$, we consider a smaller number $L_\eta \ll L$ of basis functions to prevent overfitting and reduce computational cost, resulting in a lower-rank approximation characterized by $\tilde{\Psi} \in \mathbb{R}^{V \times L_\eta}$ and $\tilde{\Lambda} \in \mathbb{R}^{L_\eta \times L_\eta}$ derived in the same manner as above, but with $L_\eta$ inducing points instead. In practice, setting $L_\eta = \lfloor 0.1L\rfloor$ is typically sufficient to capture individual spatial variation while maintaining computational efficiency and preventing overfitting.

We apply this kernel decomposition to represent the spatially varying coefficients $\bfbeta_j,\bfeta_i, \bfepsilon_i\in \mathbb{R}^V$ in vector form via the Karhunen–Loève (KL) expansion. The spatially varying components in the model can be expressed as $\bfbeta_j=\Psi\Lambda^{1/2}\bftheta_{\beta_j}$, $\bfeta_i= \tilde{\Psi}\tilde{\Lambda}^{1/2}\bftheta_{\eta_i}$, $ \bfepsilon_i=\Psi\Lambda^{1/2}\bftheta_{\epsilon_{i}}$,  where $\{\bftheta_{\beta_{j}}\}^J_{j=1}$, $\{\bftheta_{\eta_{i}}\}^N_{i=1}$, and $\{\bftheta_{\epsilon_{i}}\}^N_{i=1}$ represent the $L$-dimensional basis coefficient vectors for the population-level, individual-level, and residual spatial effects, respectively.

To allow for flexible spatially varying coefficients, we assign independent normal priors with mean zero and unknown variances for each spatial component, denoted by $\bftheta_{\beta_{j}} \sim \mathcal{N}(0, I_L\sigma^2_\beta )$, $\bftheta_{\eta_{i}} \sim \mathcal{N}(0, I_L\sigma^2_\eta )$, and $\bftheta_{\epsilon_{il}}\sim \mathcal{N}(0, I_L\sigma^2_\epsilon )$.

\section{Posterior Computation}\label{sec:post_comp}
\subsection{Model reparameterization}\label{ss:repara}
To achieve scalable Bayesian inference, we reformulate the proposed model in a low-dimensional subspace induced by the kernel approximation described in Section~\ref{ss:kernel}. Let $\bfy_i \in \mathbb{R}^V$ denote the vector of observed image responses for participant $i$, and let $\bfx_i \in \mathbb{R}^{J+1}$ be the vector of intercept and $J$ covariates. Then, Model \eqref{e:IM} can be expressed in vectorized form as:
\begin{equation}\label{e:IMP_vec}
\bfy_i = \bfx_i^\top\bfalpha \bfone_V + \bfx_i^\top \bftheta_{\beta} \Lambda^{1/2} \Psi^\top + \bftheta_{\eta_i} \tilde{\Lambda}^{1/2} \tilde{\Psi}^\top + \bfepsilon_i, \quad \bfepsilon_i \sim \mathcal{N}(0, \sigma^2_\epsilon K), \quad K \approx \Psi \Lambda \Psi^\top, 
\end{equation}
where $\bfone_V\in \mathbb{R}^V$ is a vector of all ones, $\bfalpha \in \mathbb{R}^{J+1}$ is the vector for common covariate effects shared by all locations, $\bftheta_{\beta} \in \mathbb{R}^{(J+1) \times L}$ contains the basis coefficients for the population-level effects after KL expansion, and $\bftheta_{\eta_i} \in \mathbb{R}^{L_\eta}$ contains coefficients for the participant-specific effects. Here, $\Psi \in \mathbb{R}^{V \times L}$ and $\Lambda \in \mathbb{R}^{L \times L}$ represent the approximated orthonormal basis functions and the associated eigenvalues from the low-rank kernel approximation described in Section \ref{ss:kernel}.

Direct inference in the original space remains computationally prohibitive due to the large dimension $V$. To avoid this bottleneck, we reparameterize the model by projecting the data onto the low-dimensional subspace. Specifically, we define the projection matrices:
$\Phi = \Psi \Lambda^{-1/2} \in \mathbb{R}^{V \times L}, \quad \Phi_\eta = \tilde{\Lambda}^{1/2}\tilde{ \Psi}^\top \Psi \Lambda^{-1/2} \in \mathbb{R}^{L_\eta \times L}, $
and compute the transformed response vector for participant $i$ as $\widetilde{\bfy}_i = \bfy_i^\top \Phi \in \mathbb{R}^{L}$. Under this reparameterization, the model becomes:
\begin{equation}\label{e:IMT}
\widetilde{\bfy}_i = \bfx_i^\top\bfalpha \bfone_V\Phi+ \bfx_i^\top \bftheta_{\beta} + \bftheta_{\eta_i} \Phi_\eta + \widetilde{\bfepsilon}_i, \quad \widetilde{\bfepsilon}_i \sim \mathcal{N}(0, \sigma^2_\epsilon I_L).
\end{equation}
This transformation reduces the original high-dimensional regression problem into a much lower-dimensional multivariate regression with only $L$ outcomes per participant. As a result, posterior inference requires only operations involving matrices of size $L$ or $L_\eta$, rather than $V$. This reparameterization~\citep{lin2024latent} effectively circumvents costly matrix inversions and enables efficient, vectorized posterior computations within the Bayesian framework.

The number of basis functions $L$ used in the low-rank kernel approximation is a critical hyperparameter in our model. It controls both the approximation accuracy and computational efficiency. A larger value of $L$ allows the model to capture more detailed spatial variability, which is necessary for recovering complex activation patterns in fMRI data. However, including more basis functions increases computational costs and may lead to overfitting, especially when the data contains substantial noise. Conversely, selecting too few basis functions may result in underfitting, failing to adequately represent the true underlying spatial structure and the important activation signals. As an additional practical consideration in the mechanics of identifying a reasonable basis size, a different value of $L$ defines a different approximated kernel and thereby a different likelihood function, particularly on the transformed data space. As a result, common Bayesian model selection criteria such as the widely applicable information criterion (WAIC) or the deviance information criterion (DIC) are not well-suited in our case, because the likelihood functions are not directly comparable across models with different numbers of basis functions. 

Therefore, we have developed the following approach to select the number of basis functions $L$. We first define an upper bound $L_{\max} \leq V$, which is determined by the available computational capacity. We then compute the covariance matrix $K_{L_{max}}\in\mathbb{R}^{L_{\max}\times L_{\max}}$ using the selected kernel function with hyperparameters empirically estimated from the data. From this truncated representation, we obtain the leading eigenvalues $(\lambda_l)^{L_{max}}_{l=1}$ and choose a candidate range of $L$ where the cumulative variance $\sum^L_{l=1}\lambda_l$ explains between 80\% and 98\% of the total variance $\sum^{L_{\max}}_{l=1}\lambda_l$. Within this range of $L$, we perform a grid search based on predictive performance evaluated via leave-one-out cross-validation (LOOCV). Specifically, we compute the voxel-wise predictive mean squared error (PMSE) by leaving out one participant at a time and using the fitted model from the remaining participants to predict the response. This approach allows us to select the value of $L$ that yields the best predictive accuracy without relying on likelihood comparisons. 

\subsection{Gibbs sampler}\label{ss:gibbs}
With the reparameterized model in \eqref{e:IMT} and prior specifications, all model parameters are assigned independent conjugate priors, enabling efficient posterior inference via Gibbs sampling. Specifically, all model parameters can be sampled directly from their full conditional distributions. The full set of parameters includes: $\bfTheta=\{(\alpha_j)^J_{j=0}, (\bftheta_\beta)^J_{j=0}, (\boldsymbol{\theta}_{\eta_i})_{i=1}^N, \sigma_\alpha^2, \sigma_\beta^2, \sigma_\eta^2, \sigma_\epsilon^2, a_\alpha, a_\beta, a_\eta, a_\epsilon\}$. Detailed derivations of these conditional posteriors are provided in the Supplementary Materials.

A key advantage of our formulation is that it avoids high-dimensional matrix inversions typically required in GP-based models. For models that include only an intercept term, the posterior covariance matrix of $\bftheta_\beta$ is diagonal, allowing for vectorized independent sampling of each coefficient in the $L$-dimensional vector. This structure significantly simplifies posterior computation. When additional covariates are included, the size of the posterior covariance of $\bftheta_\beta$ grows in quadratic order with the number of covariates $J$, but remains low-dimensional and computationally tractable, since in real applications $J$ is typically small. For the individual-specific basis coefficients $\bftheta_{\eta_i}$, we can consider $\Phi_\eta$ as the design matrix, and then the posterior is multivariate normal distribution with a covariance matrix in $\mathbb{R}^{L_\eta \times L_\eta}$. Since we set $L_\eta \ll L \ll V$, the dimensionality remains low, and posterior sampling remains efficient by using Cholesky decomposition on the posterior precision matrix. 

The total per-iteration computational complexity of our Gibbs sampler is $\mathcal{O}(NLL_\eta + L_\eta^3)$, with a memory cost of $\mathcal{O}(VL)$. This demonstrates a significant improvement from the original complexity of $\mathcal{O}(V^3)$ in time and $\mathcal{O}(V^2)$ in memory, which is achieved by low-rank kernel approximation and model reparameterization. The full Gibbs sampling algorithm is implemented in the Python module PyTorch~\citep{Ansel_PyTorch_2_Faster_2024} to leverage fast tensor operations and potential GPU acceleration. In practice, we run multiple MCMC chains with different initial values and assess convergence using the Gelman–Rubin diagnostic ~\citep{gelman1992inference} on conditional log-likelihoods.

\subsection{Variational Inference}\label{ss:VI}
While Gibbs sampling offers exact posterior inference under our model formulation, it can still be memory-consuming to store thousands of posterior samples with high-dimensional voxels or basis functions. To further improve computational efficiency, we in addition develop a variational inference (VI) approach as an alternative to MCMC-based methods.

Variational inference approximates the true posterior distribution by a simpler, tractable family of distributions, transforming the problem of sampling into an optimization task. Specifically, we adopt a mean-field variational family that assumes mutual independence among model parameters. The variational distribution factorizes as: $q(\bfTheta)=\prod^M_{m=1} q(\theta_m)$ where $\theta_m$ represents a parameter in $\bfTheta$.
Each variational factor has a closed-form full conditional distribution derived from the conjugate prior, as discussed in Section \ref{ss:gibbs}.

We optimize the evidence lower bound (ELBO) by performing coordinate ascent variational inference (CAVI). In each iteration, we update each variational factor in closed form while holding others fixed, using the analytical updates derived from the conjugacy of the model. This ensures that each step is computationally efficient and scalable, requiring only matrix operations involving low-dimensional quantities of size $L$ or $L_\eta$. The computational cost per iteration of the VI algorithm is comparable to that of the Gibbs sampler. However, VI typically converges in fewer iterations than MCMC, making it advantageous for large-scale applications requiring faster posterior approximations. Full derivations of variational posterior distributions are included in the Supplementary Materials.

\section{Simulation and validation}\label{s:simulation}
\subsection{Ex. 1: Simulated data}
To test the proposed model, we created synthetic datasets based on a single brain slice from a binary brain mask. These contain $4{,}999$ brain voxels and represent a realistic fMRI image spatial structure. We evaluate the performance of our proposed model under both Gibbs sampling and variational inference algorithms, compared to alternative competing methods. 

\begin{figure}[htbp]
    \centering
    \begin{subfigure}{0.8\textwidth}
    \caption{$N = 200, \sigma_\epsilon=2$}
        \centering
             \includegraphics[width=\linewidth]{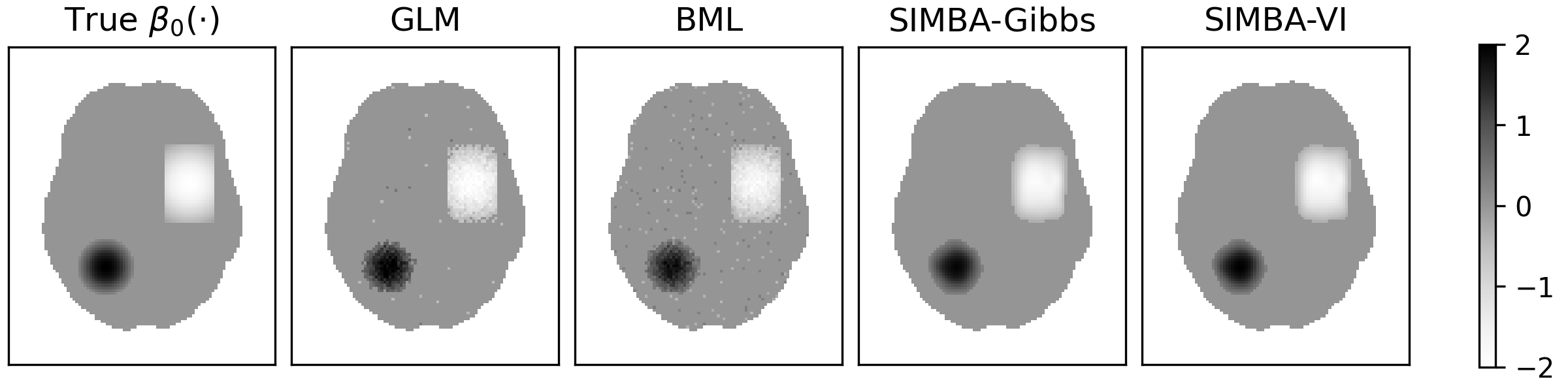}
         \label{fig:IS_res-N200_sig_eps2_beta0_0}
    \end{subfigure}
   \begin{subfigure}{0.8\textwidth}
        \centering
        \includegraphics[width=\linewidth]{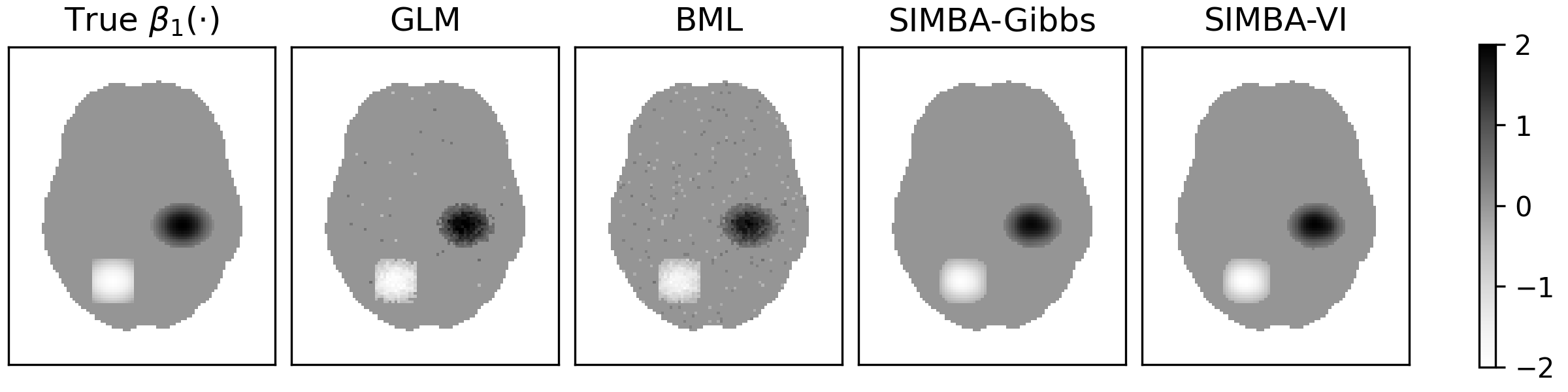}
        
         \label{fig:SI_res-N200_sig_eps2_beta1_1}
    \end{subfigure}
   
 \begin{subfigure}{0.8\textwidth}
  \caption{$N = 200, \sigma_\epsilon=5$}
        \centering
             \includegraphics[width=\linewidth]{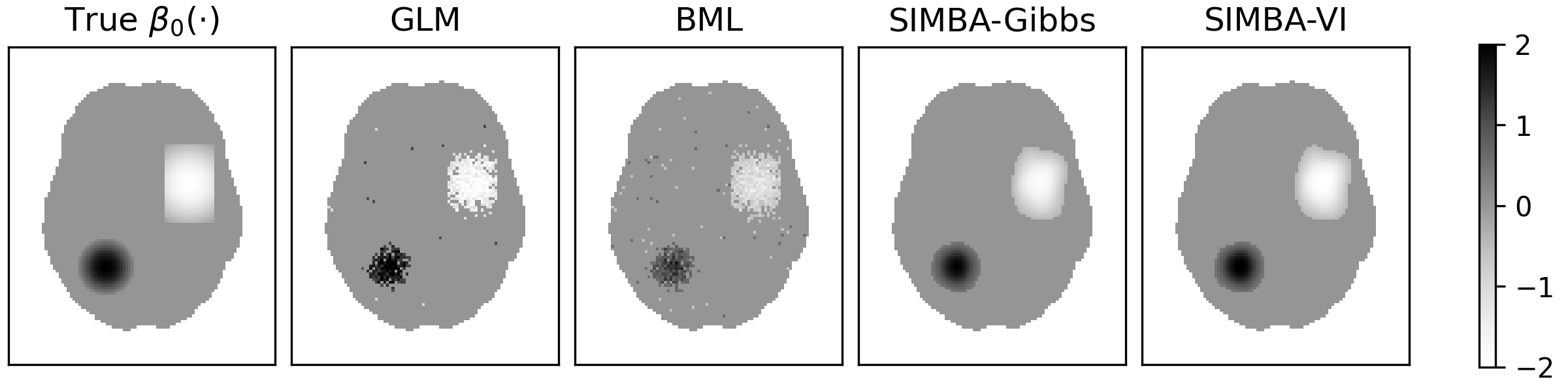}
         \label{fig:IS_res-N200_sig_eps5_beta0_49}
    \end{subfigure}

   \begin{subfigure}{0.8\textwidth}
        \centering
        \includegraphics[width=\linewidth]{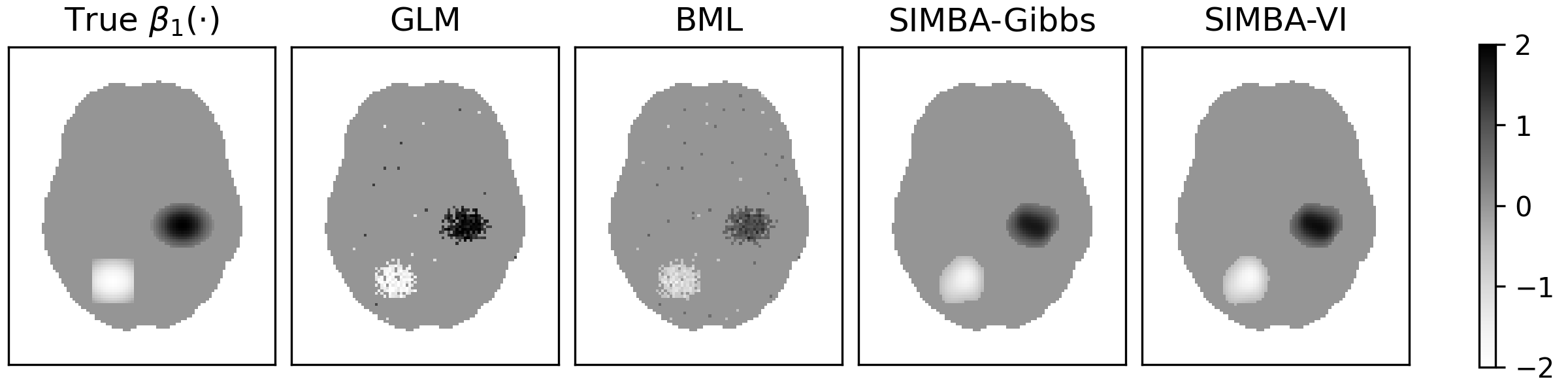}
        
         \label{fig:SI_res-N200_sig_eps5_beta1_49}
    \end{subfigure}
    \caption{ Comparison of estimated spatial effects and activation region selection across methods under two noise levels. Each row displays a map of the true spatial effect of either $\beta_0(\cdot)$ or $\beta_1(\cdot)$, as well as the corresponding estimates from each method. The non-gray shapes show the distinct "significant" activation regions, defined to be where $|E_{\rm s}|>0.95$.}
    \label{fig:sim_eff}
\end{figure}

The simulated data follow Model~\eqref{e:IM} with $J=1$ covariate. The true spatially varying intercept $\beta_0(\cdot)$ and covariate effect $\beta_1(\cdot)$ are each defined to be sparse and consist of two localized geometric regions with smoothly decaying effect sizes, as illustrated in Figure~\ref{fig:sim_eff}. In this study, we set $\alpha_0=\alpha_1=0$ to ensure comparability with competing methods. The individual-level effects $\{\eta_i(s_v)\}^n_{i=1}$ are independently drawn from a standard normal distribution. To assess model performance under different signal-to-noise ratio (SNR) conditions, we simulate datasets under two noise levels $\sigma_\epsilon=\{2, 5\}$, corresponding to SNR=$\{0.3, 0.05\}$. Under these settings, we also examine different sample sizes by considering two scenarios with number of individuals $N=\{50, 200\}$. The setting with low SNR and small sample size roughly mirrors our real fMRI data applications. Each simulation scenario is replicated with 100 datasets, each with the same fixed model parameters and different randomization seeds.

We evaluate the proposed SIMBA model using each of the Gibbs and variational inference (VI) implementations, and we also compare these to two existing approaches within the neuroimaging field. The first is the classical general linear model (GLM), which performs massive univariate regression independently at each spatial location and is a commonly used baseline in traditional fMRI analyses. This method does not incorporate spatial correlation or participant-level effects $\eta_i(s_v)$, and we use the classical Benjamini-Hochberg procedure for multiple-testing correction ~\citep{benjamini1995controlling} at FDR~=~0.05. The second comparison method is a Bayesian multi-level model (BML), which is adapted from Model (16) in \citet{chen2019handling}. Although the original formulation was designed for region-level analysis, we extend the modeling implementation to voxel-level resolution to facilitate a more direct comparison. In contrast to GLM, BML borrows strength across spatial locations through hierarchical priors, enabling partial pooling of information across both individuals and spatial locations. However, the BML method does not incorporate spatial correlations in the prior structure and models participant-level variability as constant across voxels. While the original BML model in AFNI is implemented using Hamiltonian Monte Carlo (HMC) sampling, in this voxelwise implementation we have adopted a Gibbs sampling formulation to improve computational efficiency and ensure a consistent comparison with our proposed method.

To evaluate modeling performance, we estimate the accuracy of the true spatial effects using the mean squared error (MSE) between the estimated and true main effects. In addition, we examine the activation region selection performance, which represents the ability to correctly identify the spatial locations with true non-zero effect by true positive rate (TPR) and suppress false detections in regions with no signal, as shown in the estimated FDR. For a unified assessment of significance across both frequentist and Bayesian methods, we follow the approach proposed in ~\citet{taylor2023highlight}, which converts the frequentist adjusted $p$-values and Bayesian posterior probability of positive effect $P^+=P(\beta_j(\cdot)>0)$ to a "statistical evidence" measure $E_s$ with a range of $[-1, 1]$. Specifically, for frequentist modeling, $E_s=(1-p)$ for positive effects and $E_s=-(1-p)$ for negative effects; for Bayesian modeling, $E_s=2(P^+-0.5)$ for all posterior probabilities $P^+$. For the purpose of counting true and false positives in these simulations, a voxel is referred to as "statistically significant" if $|E_s| > 0.95$, corresponding to strong evidence of a positive or negative effect. Finally, we evaluate the uncertainty quantification of each method by computing the empirical coverage rates of the true parameters, based on either confidence intervals or credible intervals, respectively.

For BML and SIMBA-Gibbs methods, we independently run three MCMC chains, each with 4,000 burn-in iterations and 1,000 posterior samples for inference. For the proposed SIMBA model, we used a Mat\'ern kernel with a length scale of 0.3, which was empirically estimated from the data. The number of basis functions $L$ used in the low-rank kernel approximation was selected for each scenario based on the lowest PMSE from LOOCV, as described in Section~\ref{ss:repara}. Here, the selected values of $L$ ranged from 90 to 160 across different simulation settings, and we set $L_\eta=\lfloor 0.1L\rfloor$ as mentioned previously to avoid overfitting and save computation costs. Empirically, we found that the estimation accuracy of the population-level main effects was not sensitive to the choice of $L_\eta$. For the simulation scenario where $N=200$ and $L=160$, the proposed SIMBA takes approximately 10 seconds per $1, 000$ iterations using the Gibbs sampler and less than 1 second using the VI algorithm, demonstrating substantial computational efficiency.

Table~\ref{tab:sim} summarizes the evaluation metrics for our proposed SIMBA model in comparison to alternative methods. Across all simulation scenarios, the SIMBA model consistently outperforms competing approaches. Specifically, SIMBA achieves the lowest mean squared error (MSE) and near-nominal coverage rates close to 95\%, illustrating accurate effect estimation. Both SIMBA implementations using Gibbs sampling and VI yield higher true positive rates (TPR) compared to the GLM and BML methods, while maintaining a low FDR below 0.05, indicating increased power for detecting true activations. In one scenario ($N=200, \sigma_\epsilon=2)$, BML achieves a slightly higher TPR, but also leads to substantially higher false signals. 

Figure~\ref{fig:sim_eff} illustrates the estimated spatial effects and selection of activation regions for all four methods, compared against the true ``significant" activation patterns (first column); each subfigure shows one representative simulation scenario from the full dataset, with a given noise level. Although the GLM and BML methods recover part of the activation regions, their estimates are noisy and spatially scattered, with notable false detections. In contrast, SIMBA more accurately reconstructs the activation regions with smooth spatially varying effects and effectively suppresses noise outside the true activation areas. The SIMBA estimates closely match the ground truth in magnitude of the effect sizes and regions with nonzero effects, demonstrating its superior ability to recover smooth spatial signals. These results demonstrate the advantages of incorporating principled spatial priors via Gaussian process modeling for voxel-wise inference, particularly in low signal-to-noise regimes.

\setlength{\tabcolsep}{8pt}
\begin{table}[htbp]
\centering
\caption{Evaluation results for simulation studies comparing mean squared error, true positive rate, false discovery rate, and coverage. Values are reported as mean percent, with standard deviations in parentheses, over 100 simulated datasets. The best-performing value(s) for each simulation is highlighted in bold. 
}
\renewcommand{\arraystretch}{1.0}
\begin{tabular*}{0.8\textwidth}{rr r rrrr}
\toprule
$N$ & $\sigma_\epsilon$ & & GLM &  BML & SIMBA-Gibbs & SIMBA-VI \\ 
\midrule
&&&\multicolumn{4}{c}{Mean Squared Error (MSE)}\\
\midrule
50  & 2 & & 9.3 (0.2) & 6.8 (0.1) & 1.2 (0.0) & \textbf{1.1} (0.0) \\
    & 5 & & 48.5 (0.8) & 14.7 (0.1) & 2.6 (0.2) & \textbf{1.9} (0.1) \\
200 & 2 & & 2.6 (0.0) & 2.3 (0.0) & 0.6 (0.0) & \textbf{0.5} (0.0) \\
    & 5 & & 13.5 (0.2) &  8.0 (0.3) & 1.4 (0.2) & \textbf{1.2} (0.1) \\
\midrule
&&&\multicolumn{4}{c}{True Positive Rate (TPR)}\\
\midrule
50  & 2 & & 61.0 (1.0) & 66.2 (1.0) & 89.0 (0.9) & \textbf{89.5} (0.9) \\
    & 5 & & 4.3 (1.0) & 1.5 (0.5) & 54.3 (2.0) & \textbf{58.2} (1.7) \\
200 & 2 & & 87.6 (0.5) &  \textbf{92.3} (0.4) & 91.9 (0.4) & 91.9 (0.5) \\
    & 5 & & 51.3 (1.1) & 58.7 (1.1) & 82.9 (0.9) & \textbf{84.1} (0.9) \\
\midrule
&&&\multicolumn{4}{c}{False Discovery Rate (FDR)}\\
\midrule
50  & 2 & & 5.1 (0.8) & 13.0 (1.0) & \textbf{6.4} (0.6) & 6.5 (0.5) \\
    & 5 & & 5.5 (3.2) & 9.9 (6.5) & \textbf{0.0} (0.0) & \textbf{0.0} (0.1) \\
200 & 2 & & 4.5 (0.6) & 20.5 (0.8) & \textbf{4.0} (0.3) & \textbf{4.0} (0.3) \\
    & 5 & & 4.5 (0.8) & 10.9 (1.0) & \textbf{2.9} (0.4) & 3.3 (0.5) \\
\midrule
&&&\multicolumn{4}{c}{Coverage }\\
\midrule
50  & 2 & & 94.6 (0.2) & 93.6 (0.2) & 97.8 (0.1) & \textbf{98.4} (0.1) \\
    & 5 & & 94.5 (0.2) & 92.0 (0.3) & 99.7 (0.2) & \textbf{100.0} (0.0) \\
200 & 2 & & 94.8 (0.2) & 94.8 (0.2) & 98.9 (0.1) & \textbf{99.1} (0.0) \\
    & 5 & & 94.9 (0.2) & 94.1 (0.3) & 98.6 (0.2) & \textbf{99.2} (0.1) \\
\bottomrule
\end{tabular*}
\label{tab:sim}
\end{table}

\section{Real data applications}\label{sec:real_data_apps}
\subsection{Ex.2: Neuroimaging Analysis Replication and Prediction Study}
We evaluate the proposed SIMBA model in a group-level analysis using publicly available task-based fMRI data from the Neuroimaging Analysis Replication and Prediction Study (NARPS) ~\citep{botvinik2020variability}. The NARPS project was a large-scale initiative designed to assess variability across neuroimaging analysis pipelines and evaluate reproducibility among research teams, which was also later re-examined separately in ~\citet{taylor2023highlight}. Both the unprocessed fMRI data and the results of the participating analysis teams have conveniently been made publicly available.

The dataset includes volumetric fMRI data collected during a mixed-gambles task, in which participants were asked to accept or reject monetary gambles with varying potential gains and losses. Each stimulus trial consists of a gain–loss pair, and participants would respond on a four-point scale (strongly/weakly accept/reject). In this study, we focus on the ``equal range'' group ($N=54$ scanned participants), where the possible gain and loss values in the trials have the same value range.

Preprocessing was conducted using standardized pipelines incorporating FreeSurfer's recon-all ~\citep{fischl2000measuring} for anatomical parcellation and AFNI's afni\_proc.py \citep{cox1996afni, reynolds_etal2024} for full participant-level processing and voxel-level contrast generation, as described in \citet{taylor2023highlight}. Participant-level contrast regression coefficients were estimated using a massive univariate time series regression model with a temporal structure of ARMA(1, 1) for the residuals at each voxel, with the gain and loss values treated as continuous amplitude modulators. The corresponding regression coefficients quantify the percentage of BOLD signal change per dollar. Specifically, we focus on the marginal effect of gain (in units of BOLD percent signal change per dollar) in this study. After participant-level processing and quality control using afni\_proc.py's QC HTML and gen\_ss\_review\_table.py (\citet{taylor_etal2024}; see \citet{taylor2023highlight} for details on criteria applied here) the final dataset included voxel-level contrasts from 47 individuals, each with $V = 222{, }230$ voxels across the brain. In this example, we consider an intercept at the group level without including additional covariates.

\begin{figure}[htbp]
    \centering
    \caption*{Comparison of Model Results in Ex. 2 (NARPS)}
 \begin{subfigure}{0.95\textwidth}
        \centering
        \caption{GLM}
            \includegraphics[width=0.95\textwidth]{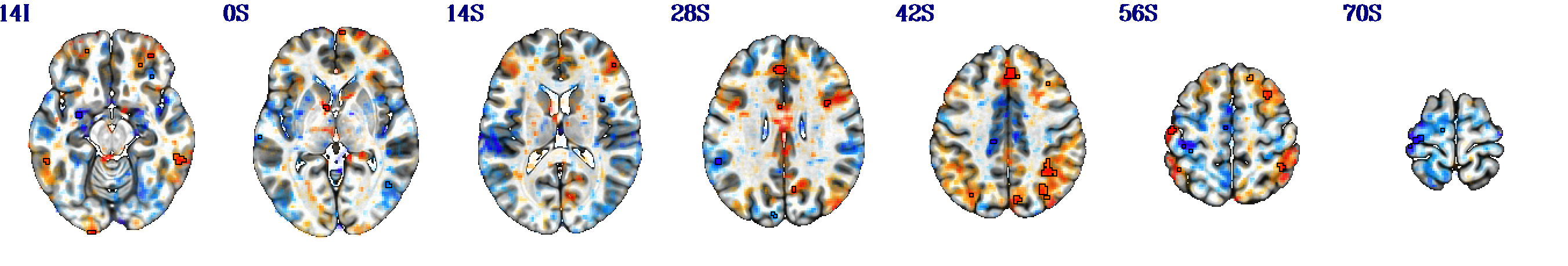}
         \label{fig:post_glm}
    \end{subfigure}

   \begin{subfigure}{0.95\textwidth}
        \centering
        \caption{BML}
        \includegraphics[width=0.95\textwidth]{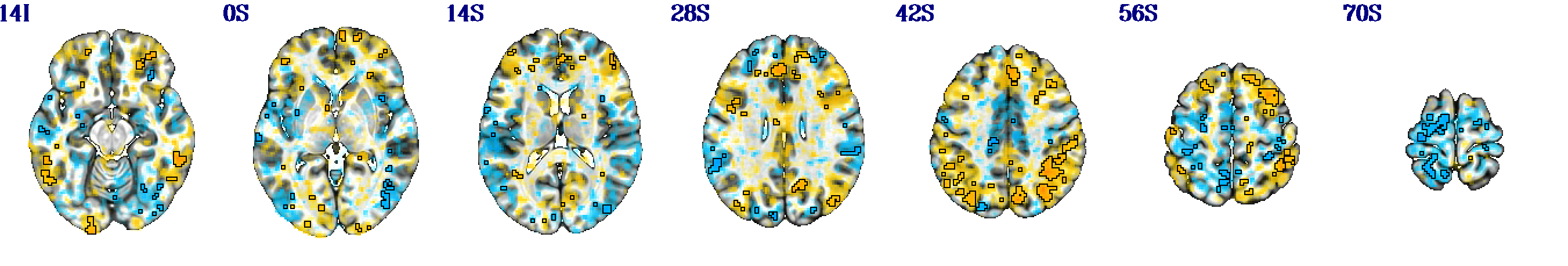}
         \label{fig:post_BML}
    \end{subfigure}

   \begin{subfigure}{0.95\textwidth}
        \centering
        \caption{SIMBA-Gibbs}
        \includegraphics[width=0.95\textwidth]{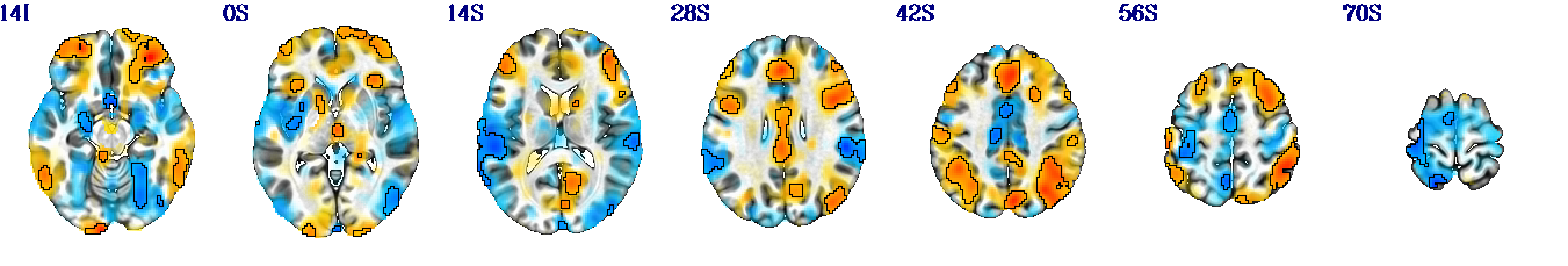}   
    \end{subfigure}
     \begin{subfigure}{0.95\textwidth}
        \centering
        \caption{SIMBA-VI}
        \includegraphics[width=0.95\textwidth]{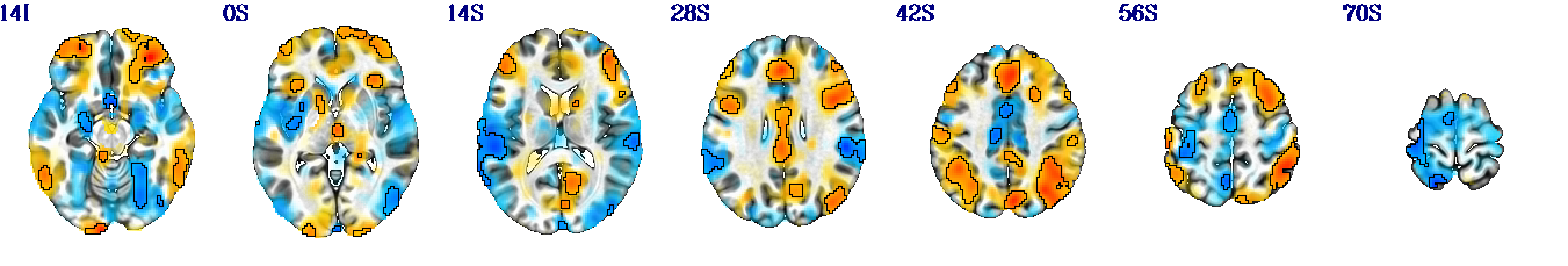}
        
         \label{fig:SI_res-narps_BSHE_VI_posterior}
    
    \end{subfigure}
    \hfill
    \begin{subfigure}{0.35\textwidth}
        \centering
        \includegraphics[width=\textwidth]{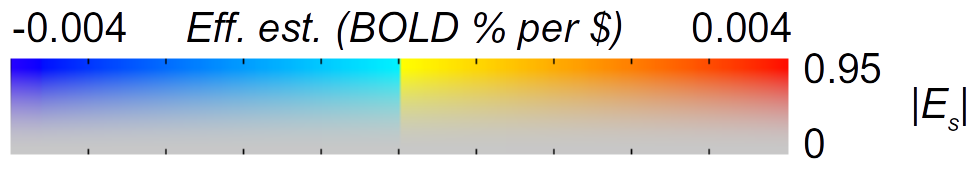}
    \end{subfigure}
    \caption{Comparison among voxel-level effect estimation in GLM, BML, and our proposed SIMBA on various axial slices of fMRI results. The overlay colors show the effect sizes, and statistical information is used to threshold the data in the "highlight" style ~\citep{allen_etal2012, taylor2023highlight}: suprathreshold regions are opaque and outlined, with other regions fading with decreasing statistical value.}
    \label{fig:post_mean}
\end{figure}

We compare our SIMBA model using both the Gibbs sampler and VI algorithm to the classical GLM and the BML models. We evaluate voxel-level effect estimates, activation detection of suprathreshold regions, and model fitting to the observed data. The activation threhsold is defined using the same statistical evidence measure $E_s$ introduced in Section~\ref{s:simulation}.  In this analysis of real fMRI data, we define the Bayesian posterior probability of positive intercept effect as $P^+=P(\alpha_0+ \beta_0(\cdot)>0)$, which includes the global intercept effect. For the GLM, voxel-wise activation is determined by $t$-tests with multiple testing adjustment at a false discovery rate of 0.05. The quality of model fit is further assessed using posterior predictive checks (PPCs) \citep{gelman1996posterior,gabry2019visualization}, by comparing predictive and observed data distributions across voxels and participants.

For the Gaussian process prior in our proposed model, we adopt a Mat\'ern kernel with a length scale of $0.1$, which was empirically estimated from the pairwise correlations of a subset of spatial locations. To implement the low-rank kernel approximation, we select $L = 1{,}200$ basis functions, chosen based on both the PMSE from LOOCV evaluation and the optimal PPC coverage of the observed data distribution. For the Bayesian models, we run three MCMC chains with different initializations, each for 5,000 iterations with 4,000 burn-in samples. 

All computations are performed on a single CPU using an Apple Silicon M3 chip with 18 GB RAM. The SIMBA model using the Gibbs sampler achieves a runtime of approximately 25 seconds per 1,000 iterations, while SIMBA's VI algorithm converges in approximately 5 seconds. 
This represents a substantial reduction in computational cost compared to the previous BML method (Model (16) in \citet{chen2019handling}), which used an HMC-based sampling algorithm and required approximately 30 minutes per 1,000 iterations for four MCMC chains, using 48 CPUs to analyze 360 brain regions (as opposed to $>$200,000 voxels, in the SIMBA cases) and 47 participants.

Figure~\ref{fig:post_mean} compares voxel-level effect estimates and activation region estimation results from the three models. Overlay colors show the effect estimates in units of BOLD percent signal change per dollar. Following the "highlight, don't hide" visualization strategy proposed by \citet{allen_etal2012} and used in \citet{taylor2023highlight}, we outline suprathreshold regions and display them with full opacity, while subthreshold areas are faded to preserve spatial context rather than being hard-thresholded. Both the GLM and BML models yield scattered selected voxels without spatial regularity, while BML show decreased effect magnitudes from the global shrinkage through shared prior variance. As shown in Figure~\ref{fig:post_mean}, the GLM approach only detects a few voxels with strong statistical evidence after FDR-based multiple testing adjustment, indicating limited statistical power. In contrast, the proposed SIMBA model obtains structured and spatially continuous patterns, suggesting improved power in detecting task-related effects and better alignment with known brain structures. Importantly, SIMBA does not artificially inflate statistical evidence, since regions with large effect sizes under GLM analysis are also captured by SIMBA, suggesting that the improvement comes from more efficient use of spatial information. This observation further highlights that FDR-based corrections in MUA may be overly conservative, discarding biologically meaningful effects. 

\begin{figure}[htbp]
    \centering
    Comparison of PPC curves in Ex. 2 (NARPS)
       \begin{subfigure}{0.35\textwidth}
        \centering
\includegraphics[width=0.95\textwidth]{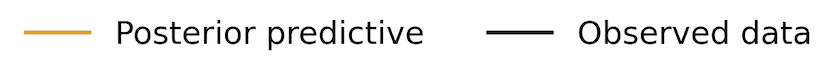}
         \label{fig:post_glm}
    \end{subfigure}\\
    \begin{subfigure}{0.24\textwidth}
        \centering            \includegraphics[width=\textwidth]{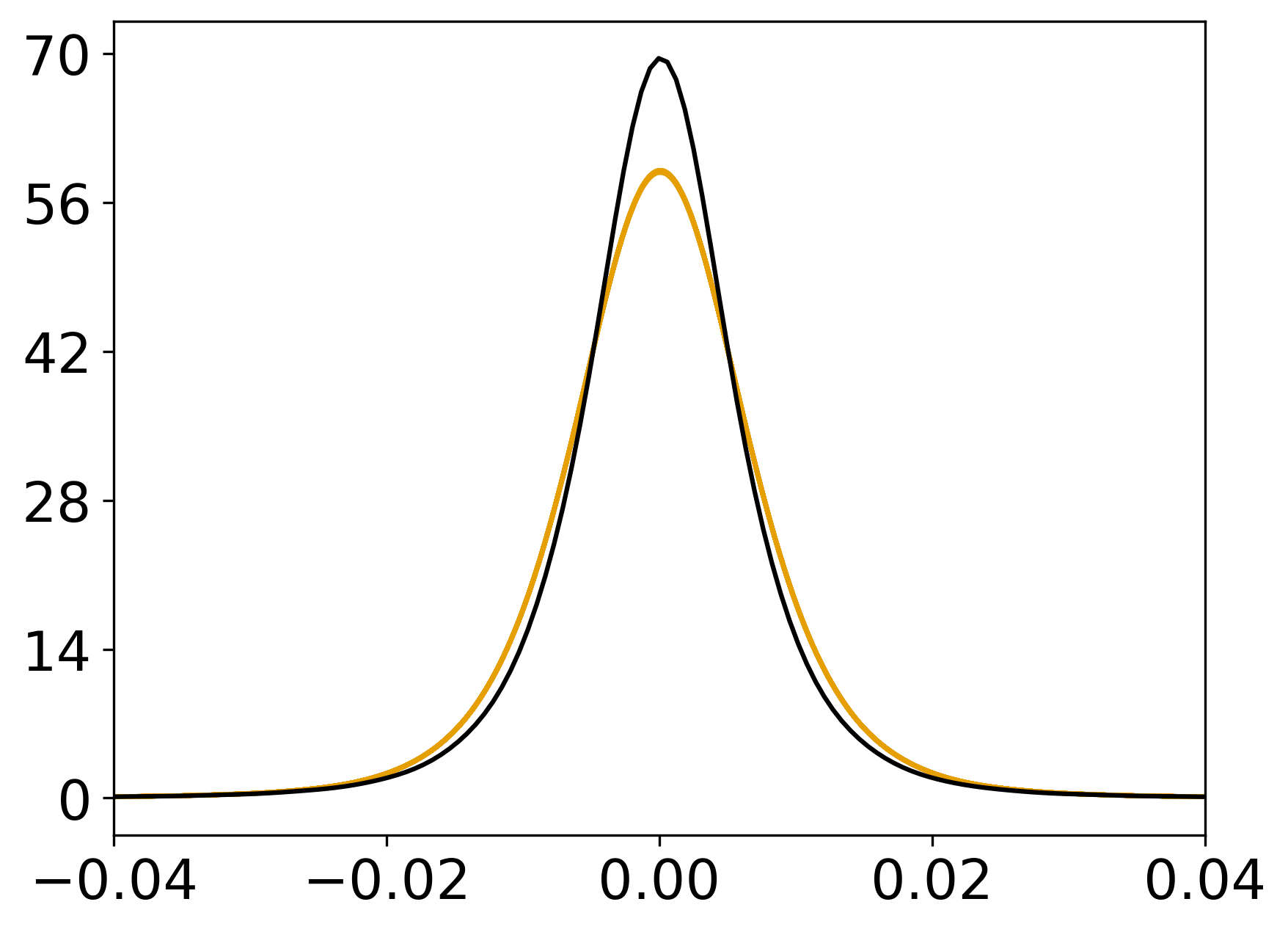}
        \caption{GLM}
         \label{fig:IS_res-ttest_ppc}
    \end{subfigure}
\hfill
 \begin{subfigure}{0.24\textwidth}
        \centering
            \includegraphics[width=\textwidth]{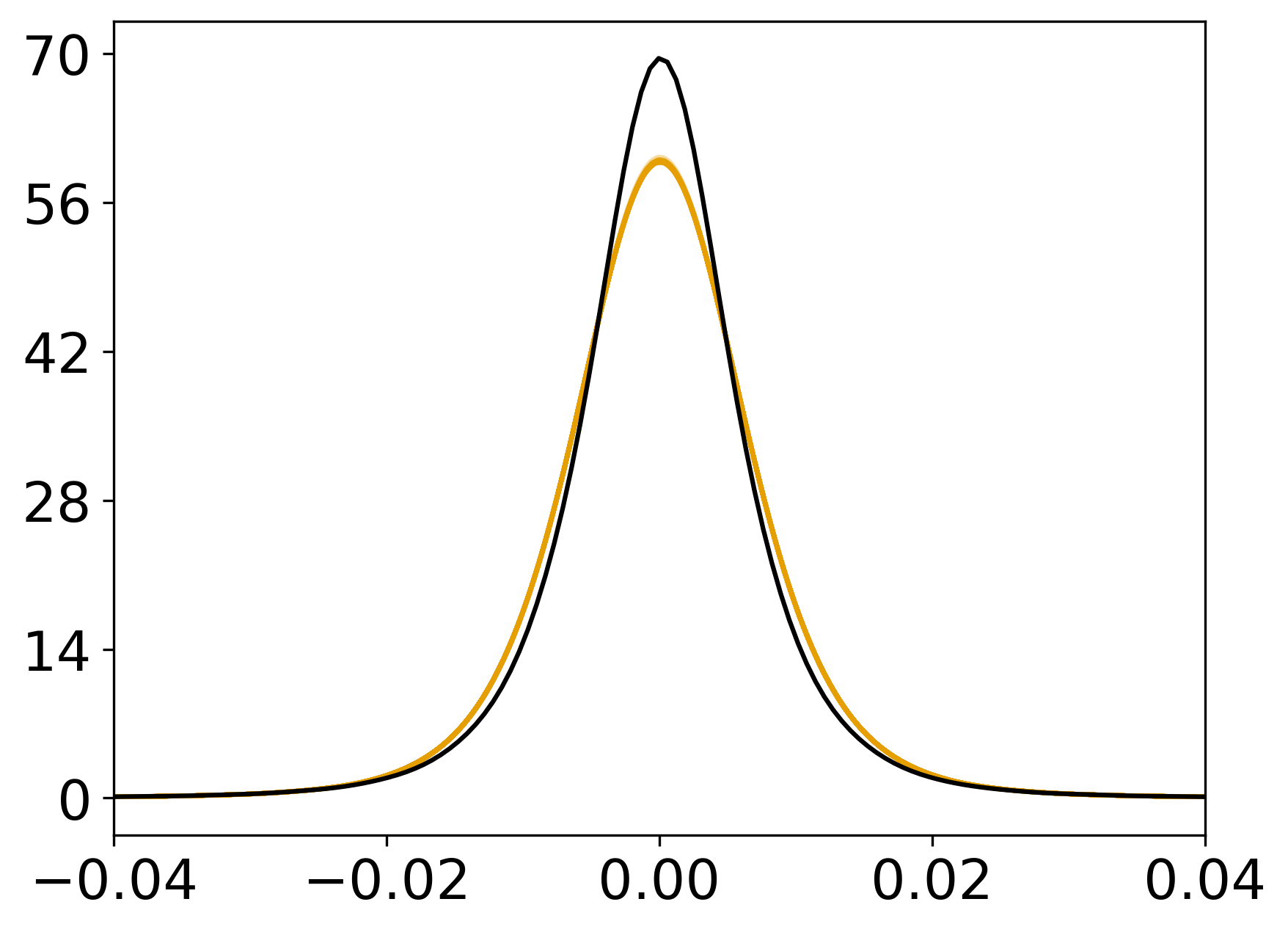}
        \caption{BML}
         \label{fig:IS_res-M1_ppc}
    \end{subfigure}
\hfill
   \begin{subfigure}{0.24\textwidth}
        \centering
        \includegraphics[width=\textwidth]{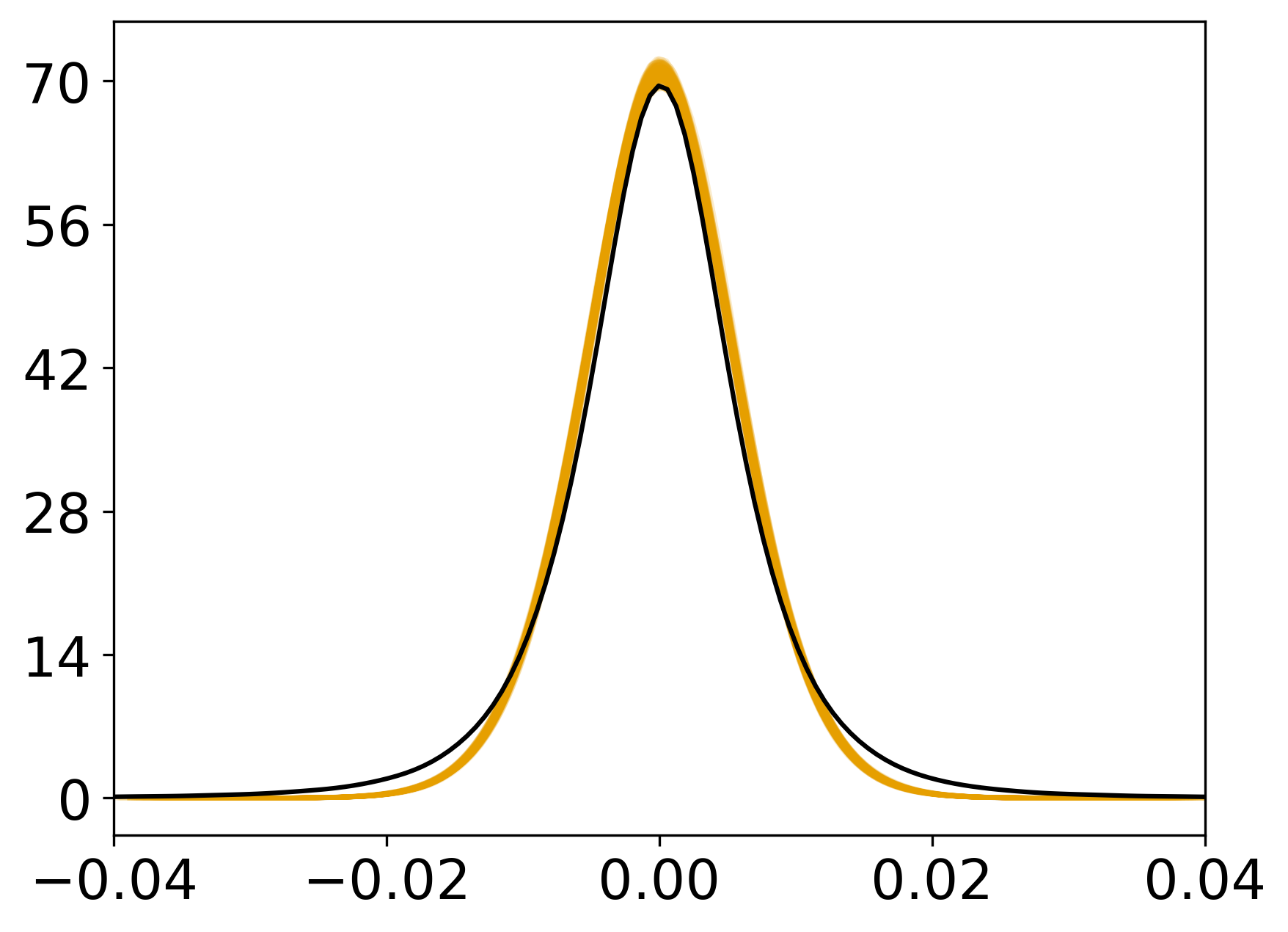}
        \caption{SIMBA-Gibbs}
         \label{fig:SI_res-M21_ppc_L1200}
    \end{subfigure}
\hfill
     \begin{subfigure}{0.24\textwidth}
        \centering
        \includegraphics[width=\textwidth]{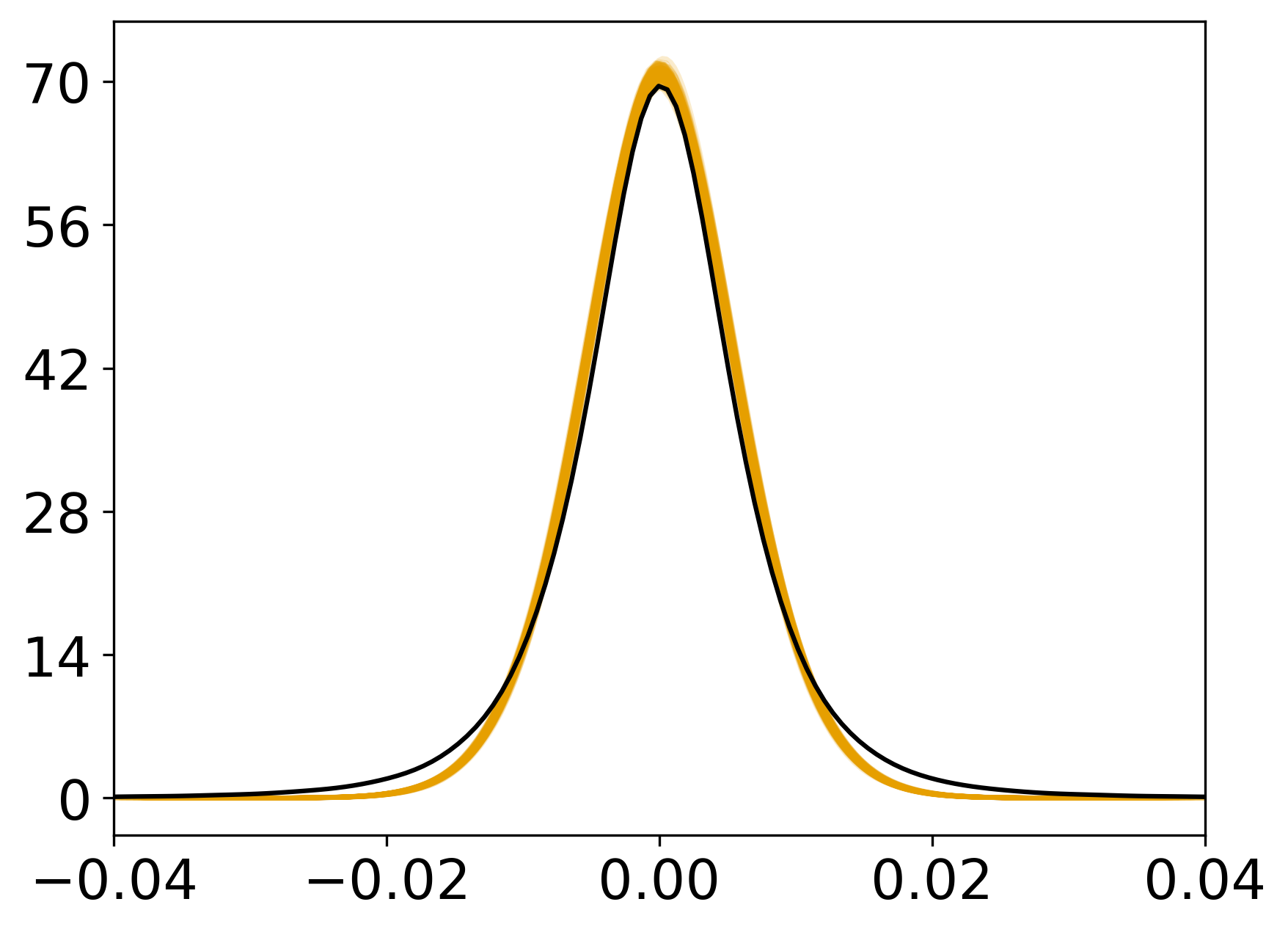}
        \caption{SIMBA-VI}
         \label{fig:SI_res-SIMBA_VI_ppc_J1200}
    \end{subfigure}
    \caption{Posterior predictive checking curves for real fMRI data (Ex. 2, NARPS). The yellow lines represent 150 posterior draws of predictive data distribution and the black line shows the observed data density for all voxels. Having a larger overlap and centering of posterior draws over the observed data fit (black line) suggests better model fit. Here, the SIMBA models have much better posterior check results than the other two models.}
    \label{fig:ppc}
\end{figure}

Finally, the PPC curves provide an important quantitative validation of our proposed framework. Figure~\ref{fig:ppc} shows 150 posterior draws of predictive data distributions (yellow lines) along with the observed data density (black line) for each method. The closely matching distributions in Figure~\ref{fig:SI_res-M21_ppc_L1200} and~\ref{fig:SI_res-SIMBA_VI_ppc_J1200} indicate that SIMBA better captures the distribution features of the data than the competing approaches. This shows the ability of the SIMBA model to learn the complex spatial characteristics of fMRI data via the Gaussian process prior.

\subsection{Ex. 3: Adolescent Brain Cognitive Development Study}
To further evaluate the efficacy of our proposed model, we additionally apply SIMBA to analyze the relationships between cortical gray matter surface fMRI data and clinical covariates from the Adolescent Brain Cognitive Development (ABCD) study in Release 4.0. The ABCD study is a large, multi-site longitudinal initiative aimed at understanding brain development from childhood through adolescence and identifying its biological and environmental determinants ~\citep{casey2018adolescent}. In this case, 
we provide an example of investigating the relationship between cognitive function and brain activity during working memory tasks using $n$-back cortical surface fMRI data. Specifically, we analyze the 2-back versus 0-back contrast results (which are measured in units of $z$-statistics), derived from the fMRI data collected at 21 study sites in the baseline study cohort.

The cortical surface data are stored on two-dimensional meshes, which contain $V=29{,}696$ and $V=29{,}716$ spatial locations for the left and right hemispheres, respectively. In the present analysis, we represent these vertices using their three-dimensional geometric coordinates to capture spatial positions in the brain, but we do not explicitly incorporate surface topology or mesh adjacency into the modeling framework (see the Discussion for future extensions with surface datasets). The processed dataset includes $N=2{,}497$ individuals with complete imaging and covariate information for analysis. Our group-level model specification\footnote{Note that our primary interest in the present work is testing the model approach with a realistic setup of fMRI data; to address specific questions of detailed brain function and cognitive ability in practice, one would need to perform a more rigorous procedure for covariate selection, such as through a causal inference framework.} includes one spatially varying intercept along with $J=4$ covariates: the general cognitive ability component score ~\citep{sripada2020prediction}, which is our primary predictor, and the psychiatric diagnostic score, age, and sex. For BML and SIMBA-Gibbs models, we ran 4,000 burn-in iterations with 1,000 posterior samples in three MCMC chains. The SIMBA model takes approximately 14 minutes for Gibbs sampling and 1.4 minutes for the VI algorithm to converge.

\begin{figure}[htbp]
    \centering
    \caption*{Comparison of Cross-Site PMSEs in Real Data (Ex. 3: ABCD)}
    \begin{subfigure}{0.48\textwidth}
        \centering
            \includegraphics[width=\textwidth]{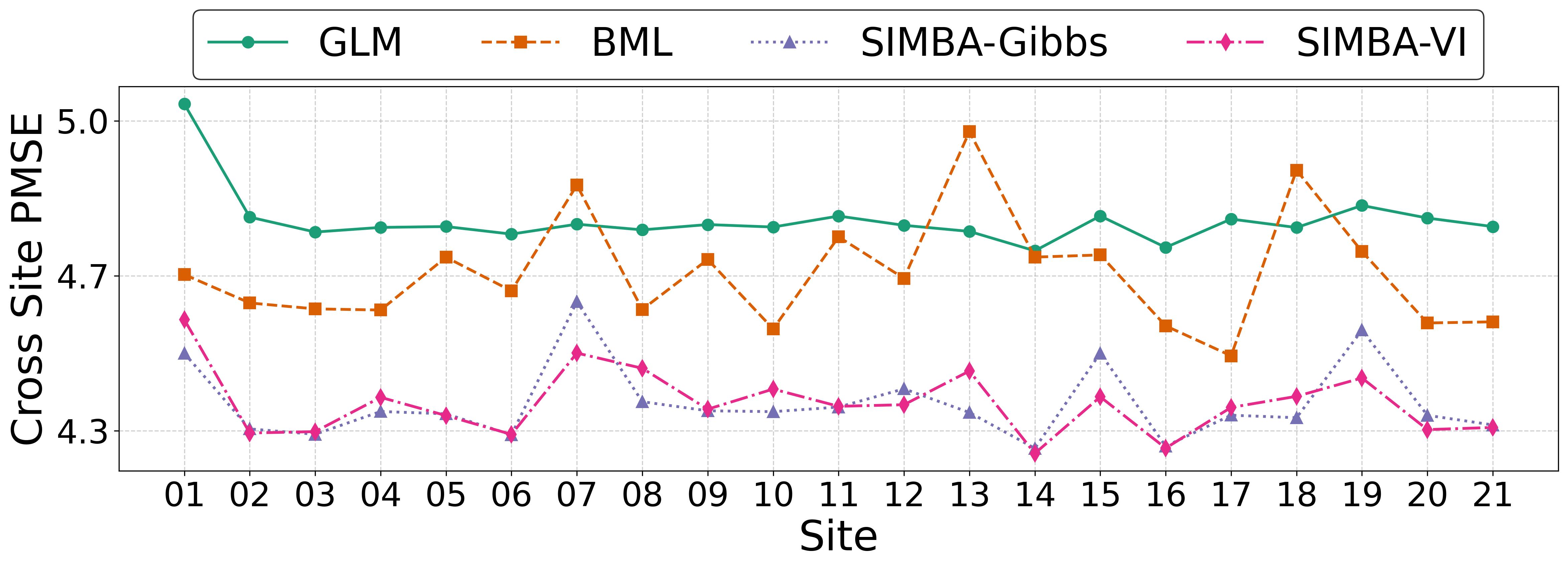}
        \caption{Left hemisphere}
         \label{fig:IS_res-left_mse_plot2}
    \end{subfigure}
\hfill
 \begin{subfigure}{0.48\textwidth}
        \centering
            \includegraphics[width=\textwidth]{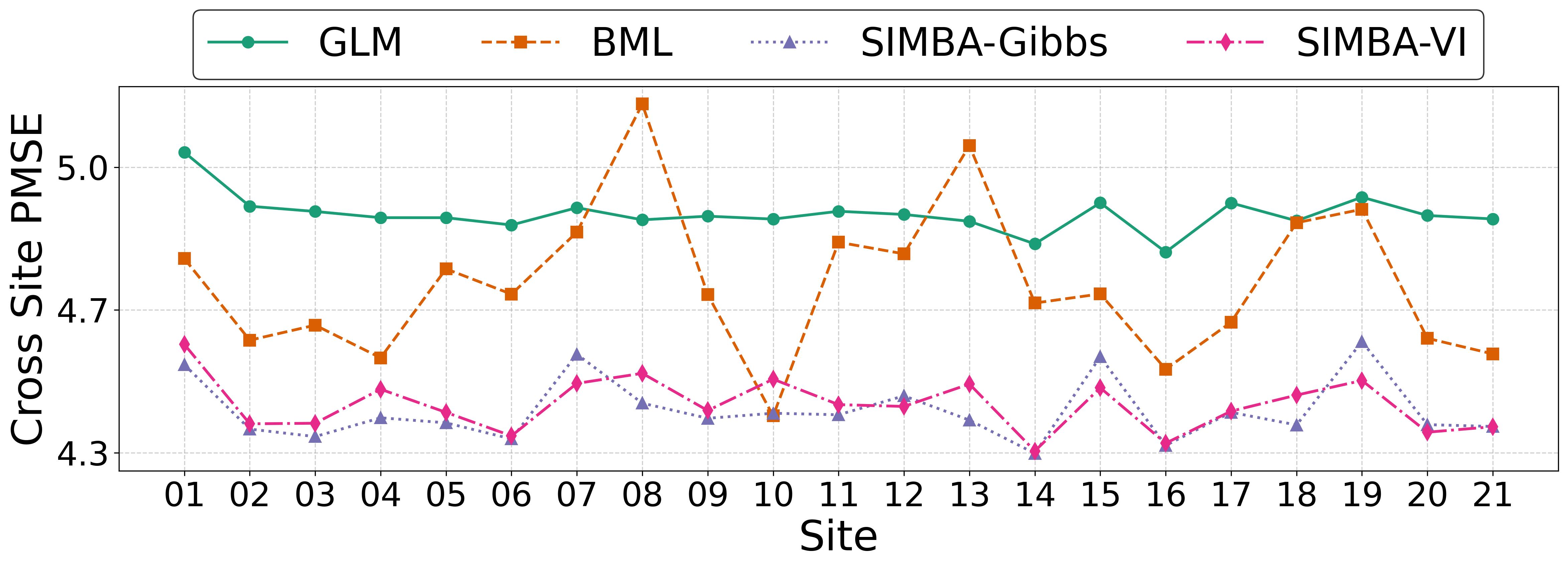}
        \caption{Right hemisphere}
         \label{fig:IS_res-right_mse_plot3}
    \end{subfigure}
    \caption{Cross-site prediction mean squared errors (MSEs) for single-site analyses on the ABCD cortical surface fMRI data, shown separately for the left hemisphere (left) and right hemisphere (right).}
    \label{fig:ABCD_MSE}
\end{figure}

\begin{figure}[htbp]
    \centering
    \caption*{Comparison of PPC curves on left (top) and right (bottom) hemisphere in Ex. 3 (ABCD)}
       \begin{subfigure}{0.35\textwidth}
        \centering
\includegraphics[width=0.95\textwidth]{figures/ppc_leg.png}
    \end{subfigure}\\
    \begin{subfigure}{0.24\textwidth}
        \centering            \includegraphics[width=\textwidth]{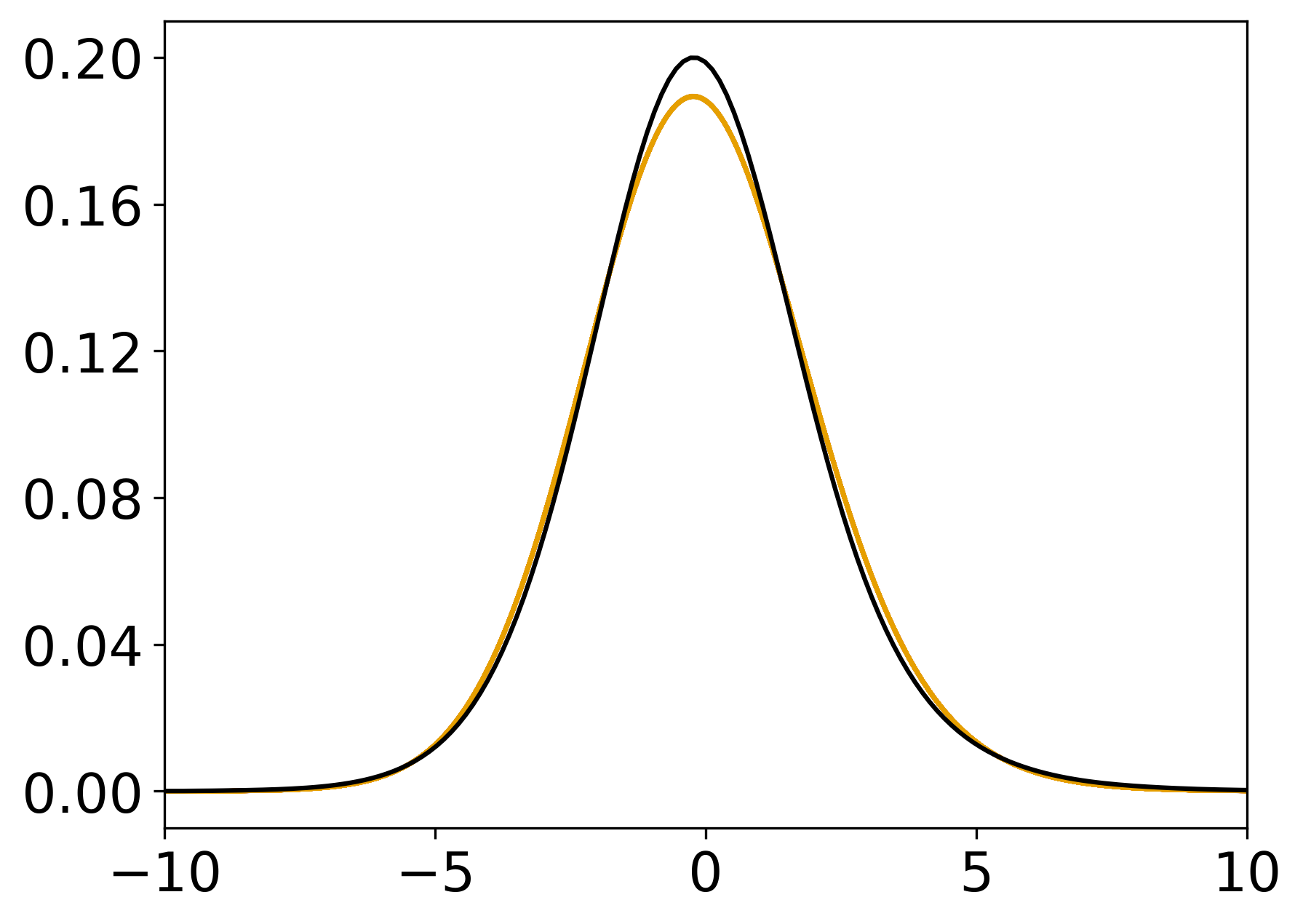}
        \includegraphics[width=\textwidth]{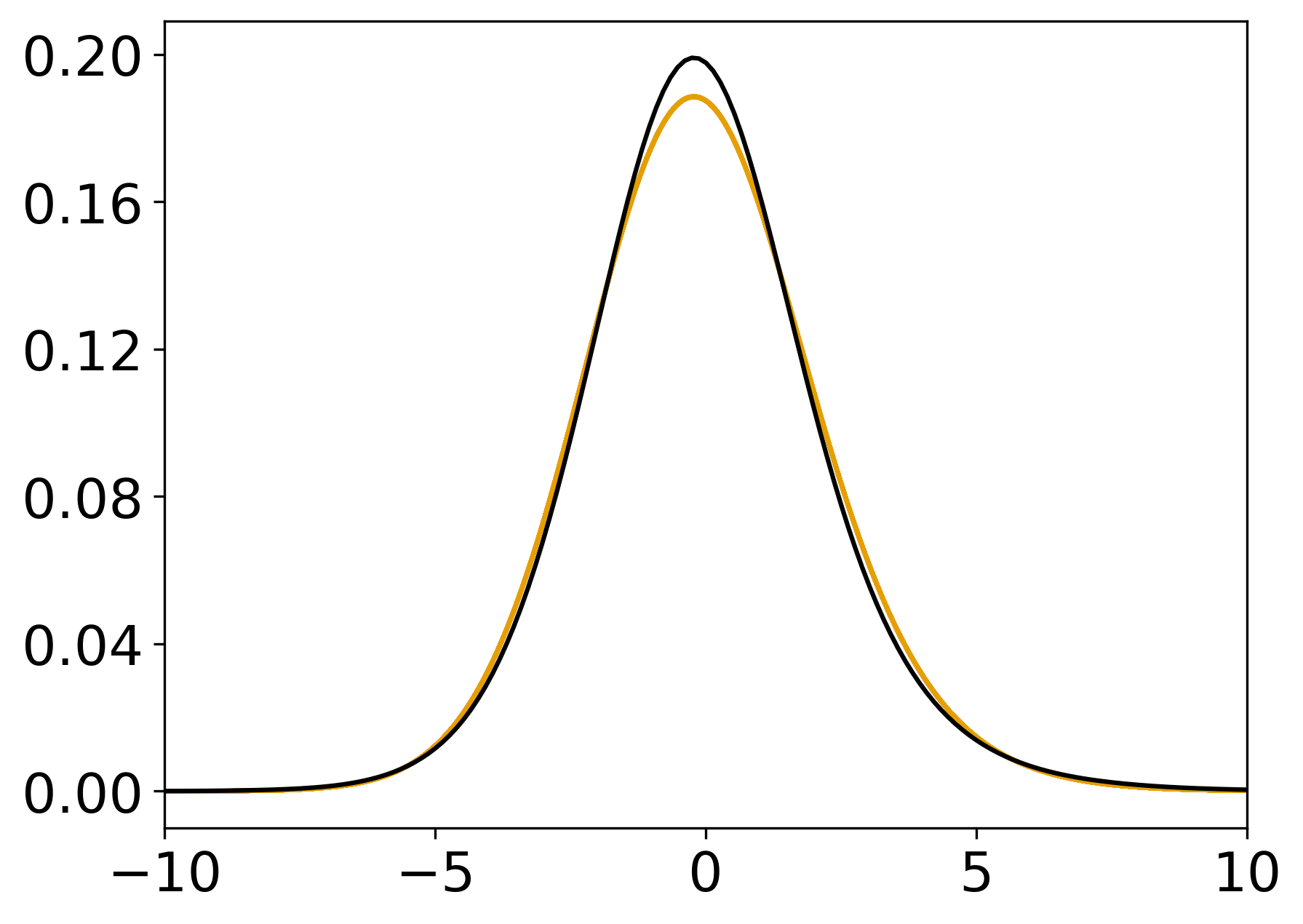}
        \caption{GLM}
         \label{fig:abcd-ttest_ppc}
    \end{subfigure}
\hfill
 \begin{subfigure}{0.24\textwidth}
        \centering
            \includegraphics[width=\textwidth]{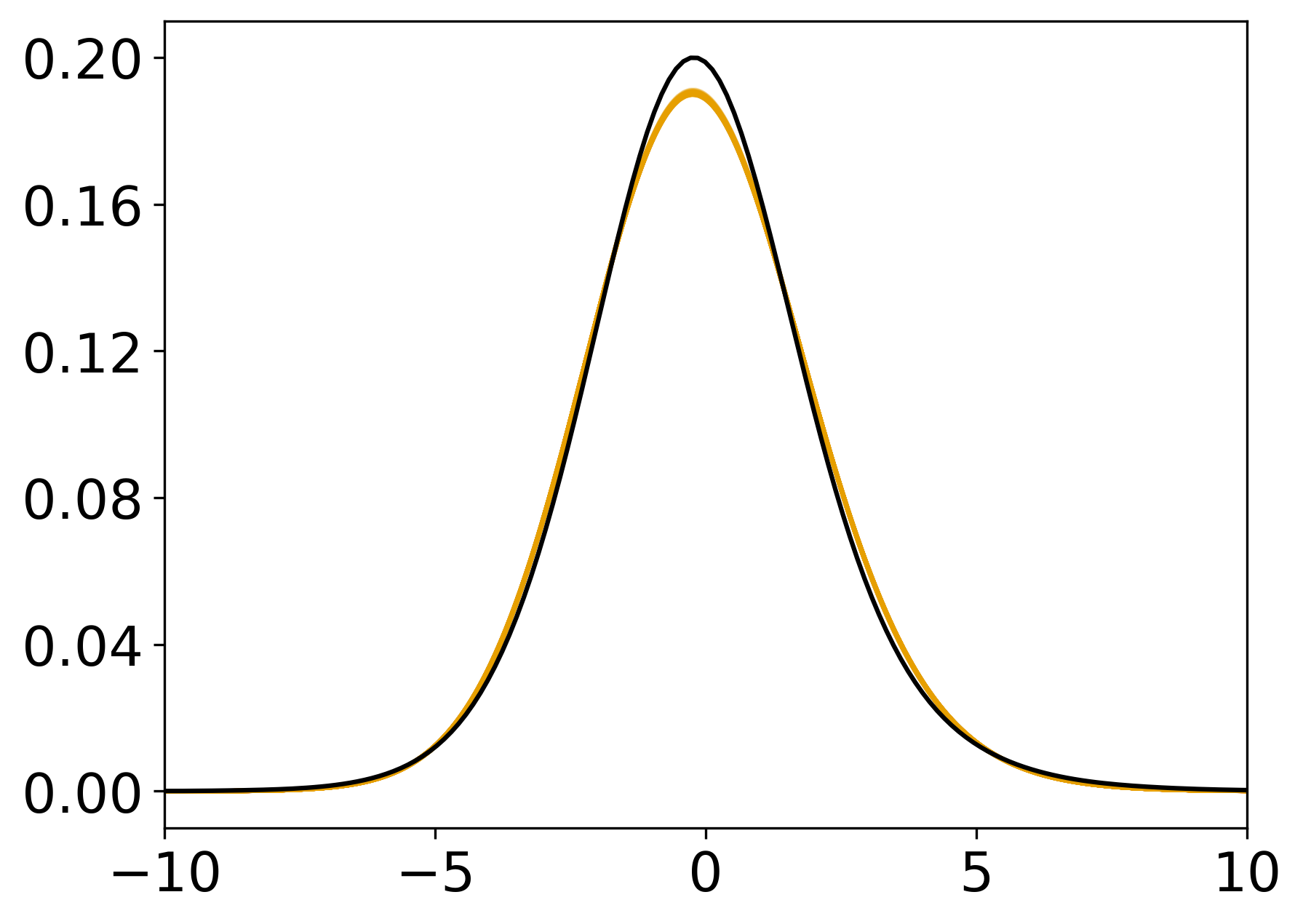}
              \includegraphics[width=\textwidth]{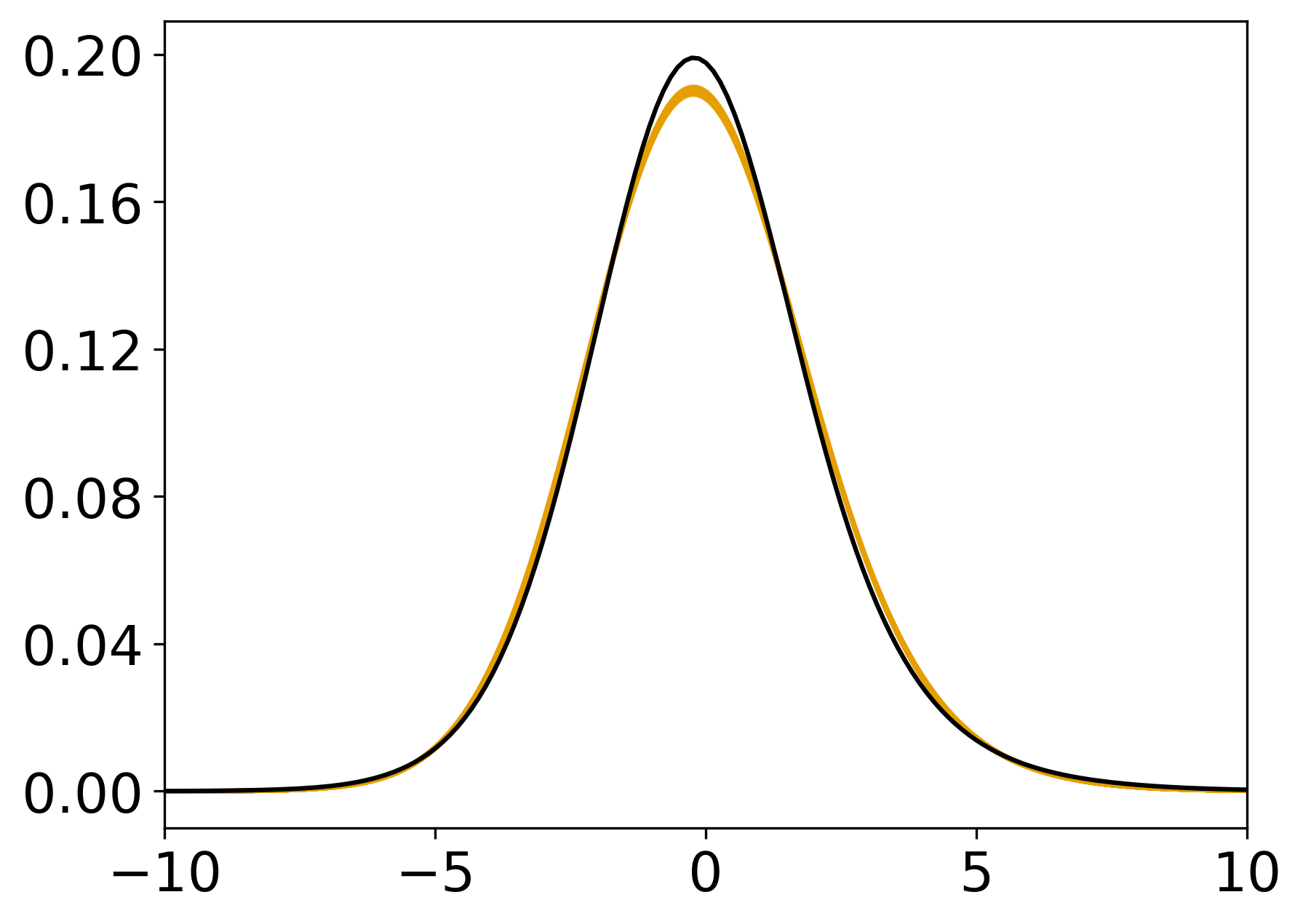}
        \caption{BML}
         \label{fig:abcd-BML_ppc}
    \end{subfigure}
\hfill
   \begin{subfigure}{0.24\textwidth}
        \centering
        \includegraphics[width=\textwidth]{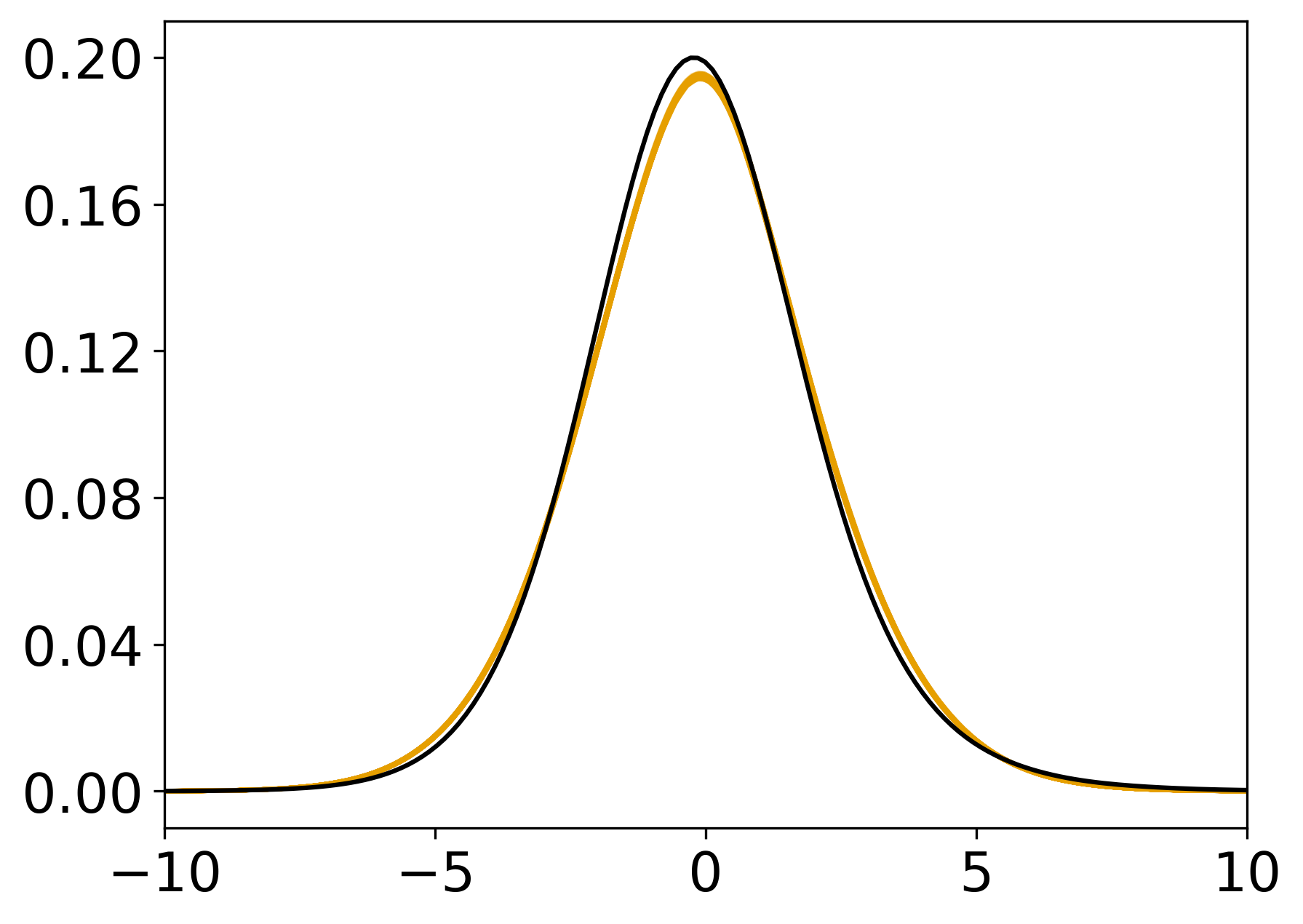}
          \includegraphics[width=\textwidth]{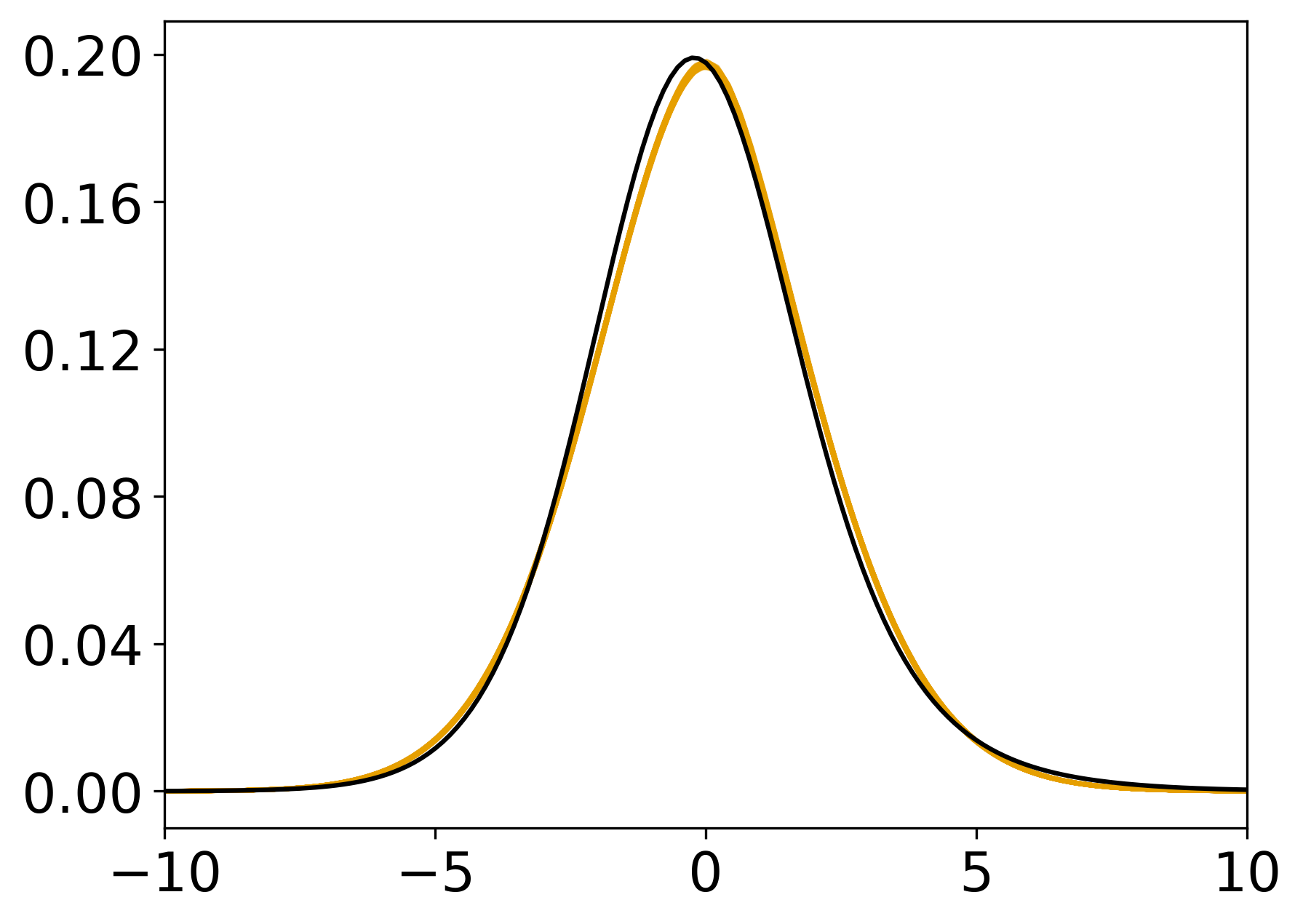}
        \caption{SIMBA-Gibbs}
         \label{fig:abcd-SIMBA_gibbs_ppc}
    \end{subfigure}
\hfill
     \begin{subfigure}{0.24\textwidth}
        \centering
        \includegraphics[width=\textwidth]{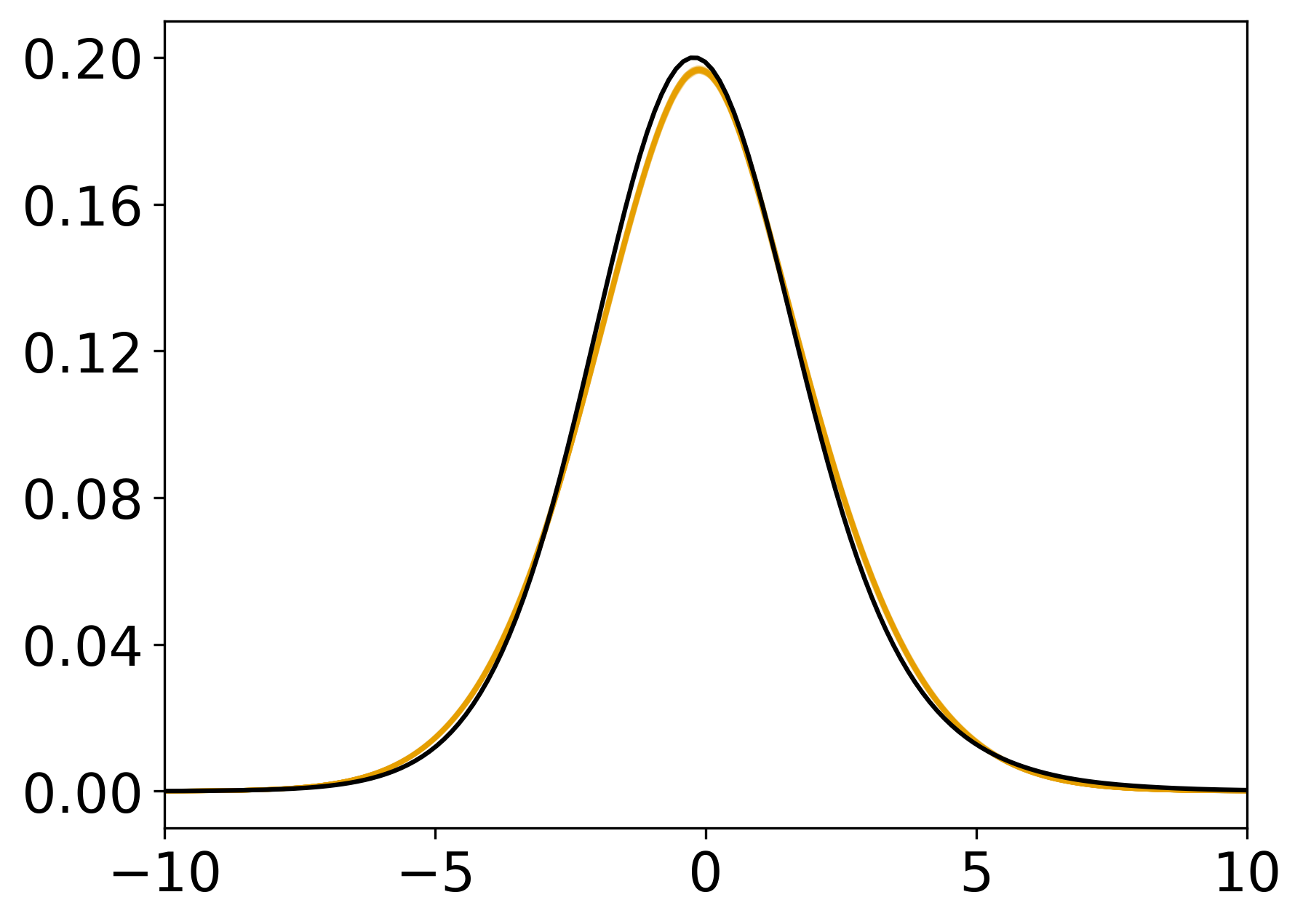}
          \includegraphics[width=\textwidth]{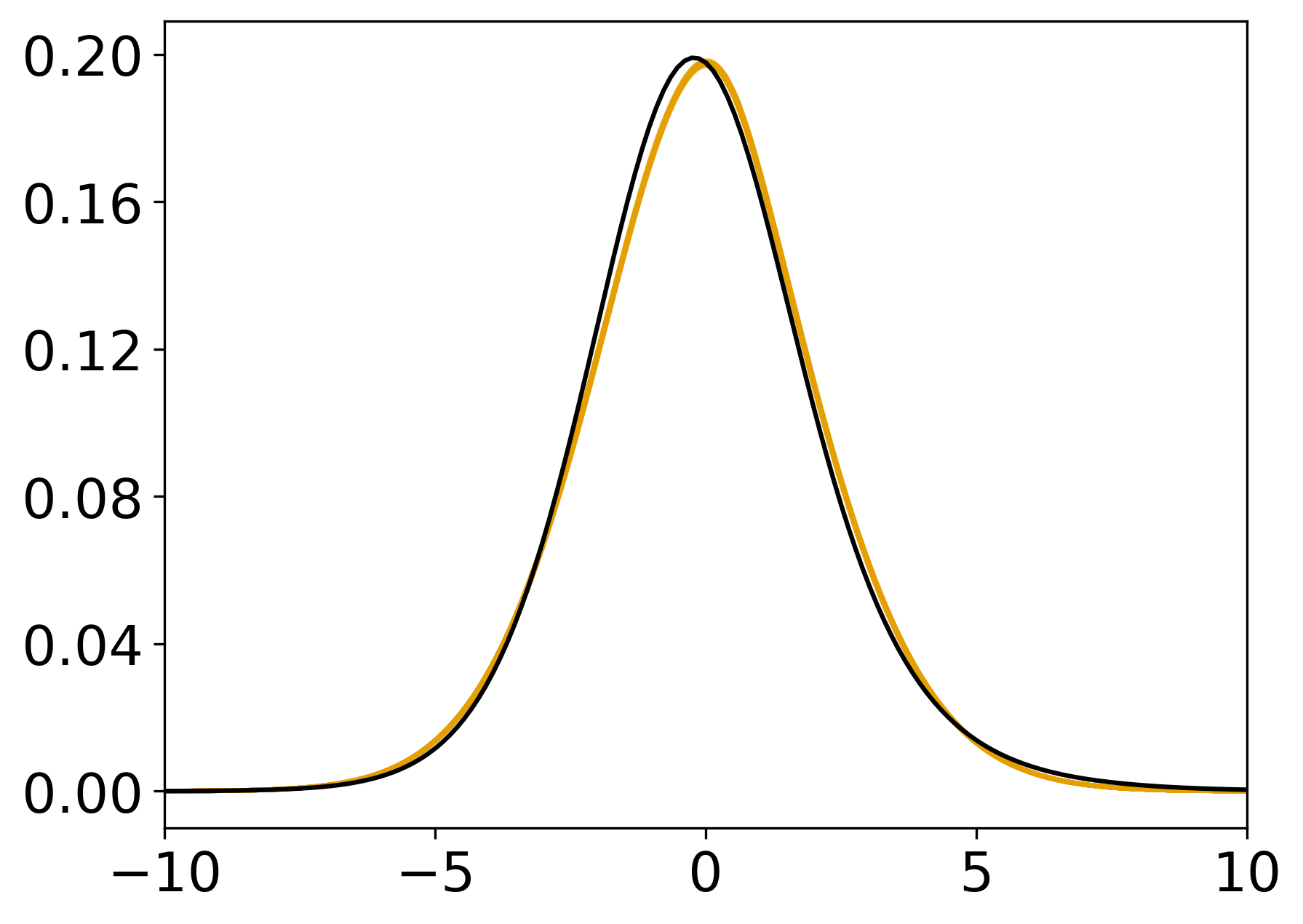}
        \caption{SIMBA-VI}
         \label{fig:abcd-SIMBA_VI_ppc}
    \end{subfigure}
    \caption{Posterior predictive checking curves for real fMRI data (Ex. 3, ABCD) on left (top row) and right (bottom row) hemispheres. The yellow lines represent 100 posterior draws of predictive data distribution, and the black line shows the observed data density for all voxels. Having a larger overlap and centering of posterior draws over the observed data fit (black line) suggests better model fit. Here, the SIMBA models have better posterior check results than the other two models.}
    \label{fig:abcd_ppc}
\end{figure}

In multi-site neuroimaging studies, scanner differences, acquisition protocols, and site-specific demographics can introduce variability that challenges model generalization. To evaluate whether the model produces stable and accurate estimates that can be generalized across study sites, we adopt a cross-site prediction framework. For each site, we fit the model using only its data, then predict imaging outcomes for participants from all other sites. We evaluate the performance by the out-of-site prediction MSEs as an external validation metric. Figure~\ref{fig:ABCD_MSE} presents the cross-site prediction MSEs for both hemispheres among the baseline cohort. Across both hemispheres, the GLM typically shows the highest prediction MSE and the largest variability across sites, indicating limited generalizability. BML mostly performs better than GLM, but it is generally outperformed by our proposed SIMBA methods. Both SIMBA models with Gibbs sampler and VI implementations achieve consistently lower or comparable MSEs in nearly all sites. These results demonstrate that SIMBA better captures the spatial structure to achieve more accurate and stable estimation results and cross-site predictions, highlighting its robustness to site heterogeneity. Similar to Ex.~2, we compare the posterior model fitting on both left and right hemispheres by examining the corresponding PPC curves. Figure~\ref{fig:abcd_ppc} shows 100 posterior draws of predictive data distributions (yellow lines) along with the observed data density (black line) for each method. The matching distributions in Figure~\ref{fig:abcd-SIMBA_gibbs_ppc} and~\ref{fig:abcd-SIMBA_VI_ppc} illustrate that both Gibbs and VI implementations in SIMBA more accurately capture data variability than the GLM and BML counterparts, further highlighting the robustness of SIMBA in modeling whole-brain spatial patterns of regression effects in large-scale fMRI data.

\begin{figure}
[htbp]
    \centering
    \caption*{Comparison of Model Results in Ex. 3 (ABCD)}
   
    \begin{tabular}{c ccccr}
    & GLM & BML & SIMBA-Gibbs & SIMBA-VI & \\ \toprule
  \raisebox{10mm}{Left} & \includegraphics[width=0.15\textwidth]{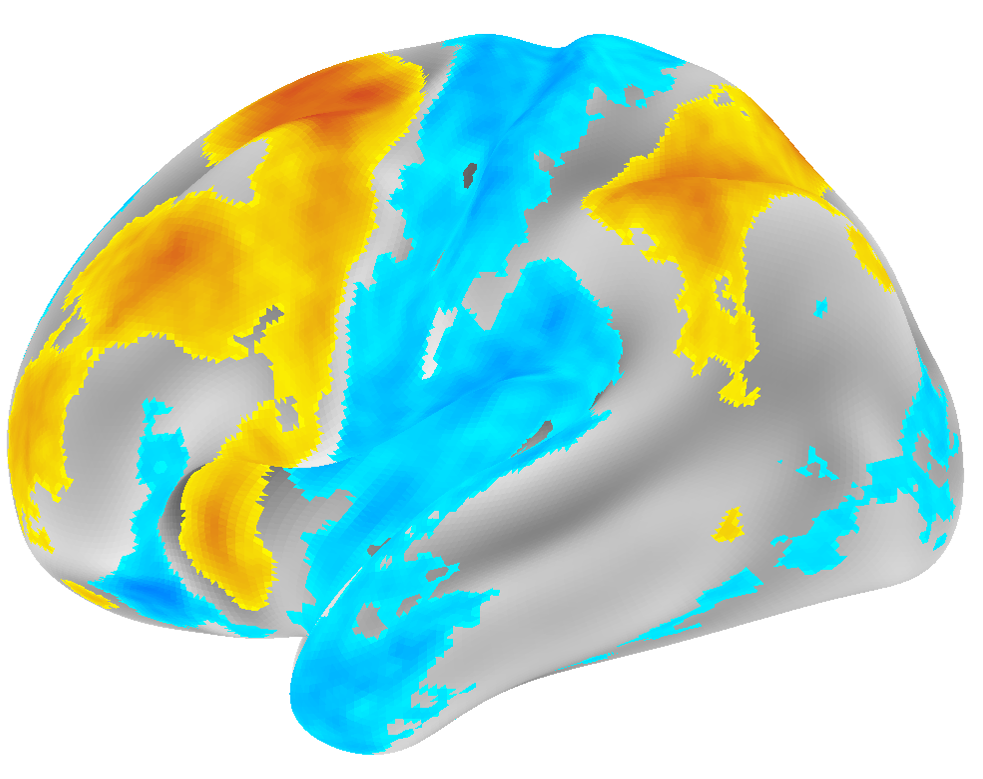} & \includegraphics[width=0.15\textwidth]{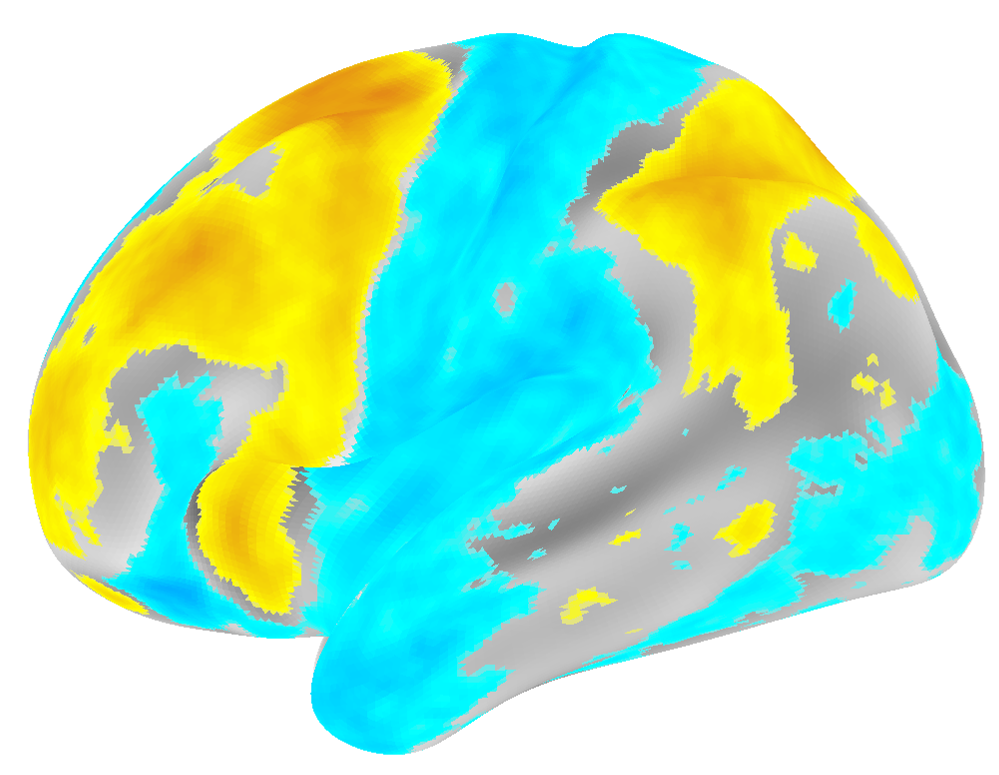} & \includegraphics[width=0.15\textwidth]{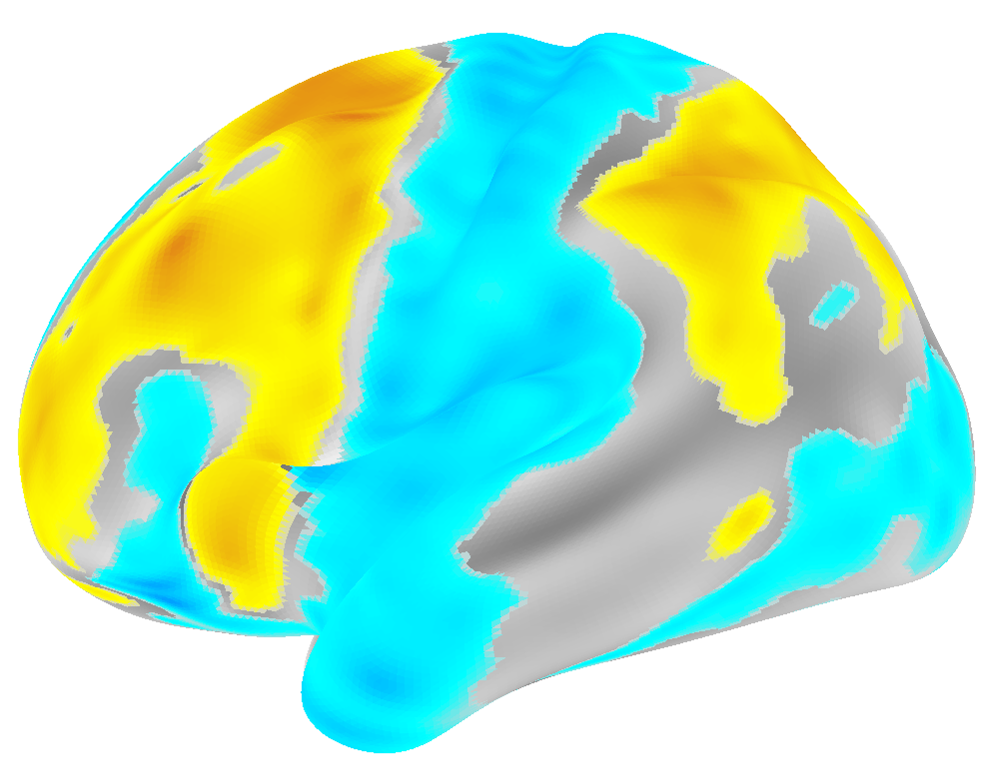} & \includegraphics[width=0.15\textwidth]{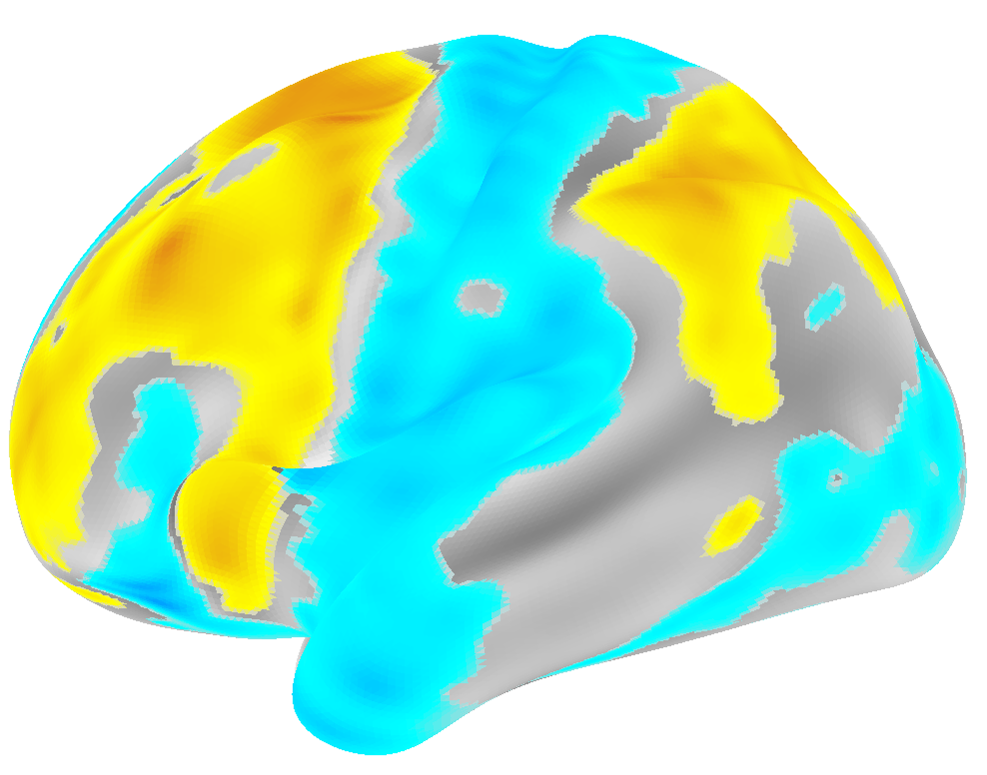} &
  \raisebox{15mm}{
  \multirow{2}{*}{
\includegraphics[width=0.13\linewidth]{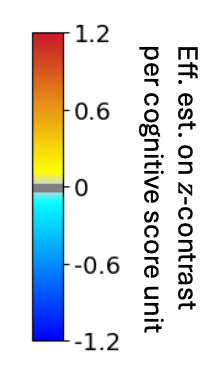}}}\\
  \raisebox{10mm}{Right} & 
  \includegraphics[width=0.15\textwidth]{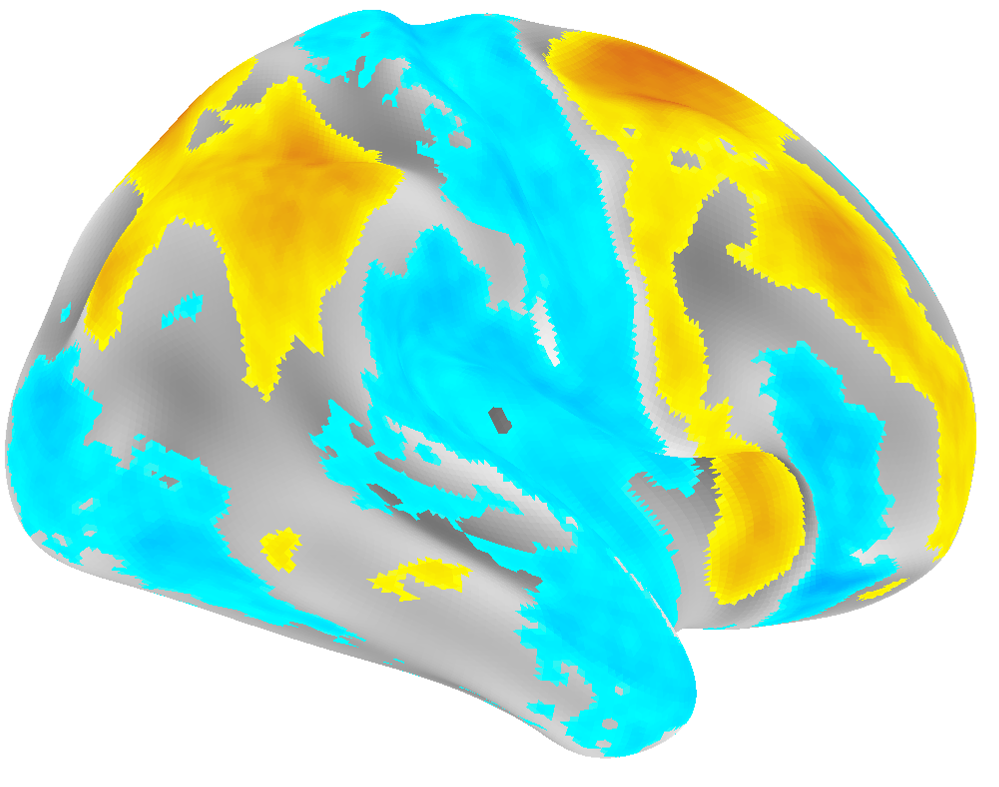} & \includegraphics[width=0.15\textwidth]{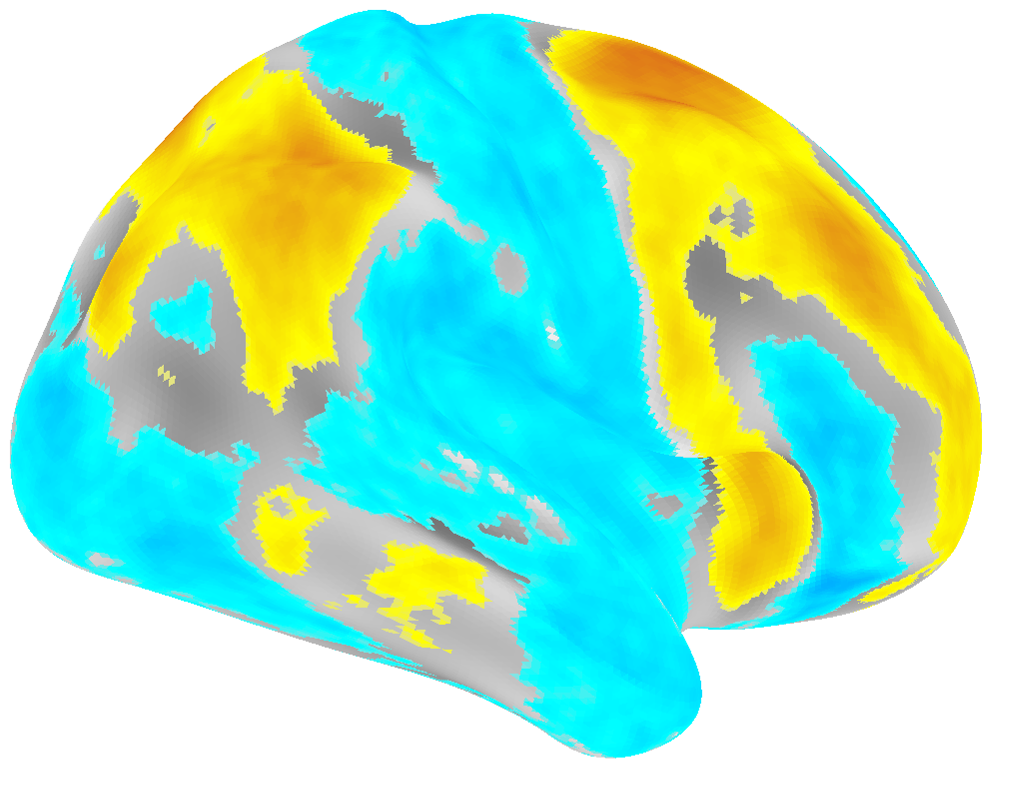} & \includegraphics[width=0.15\textwidth]{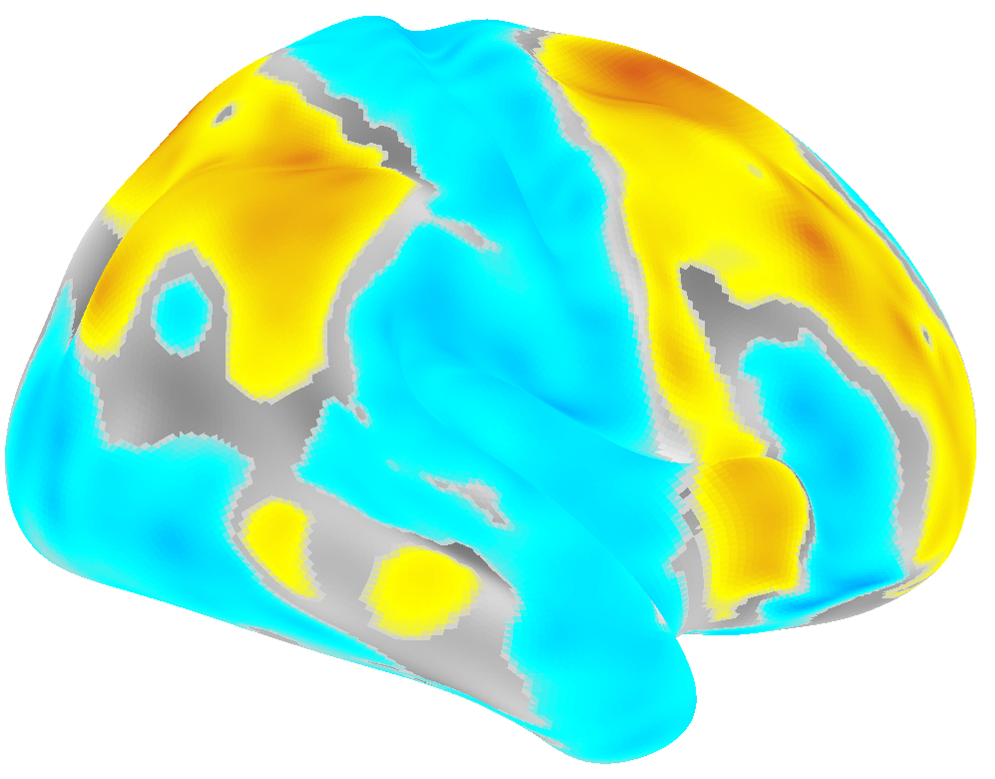} & \includegraphics[width=0.15\textwidth]{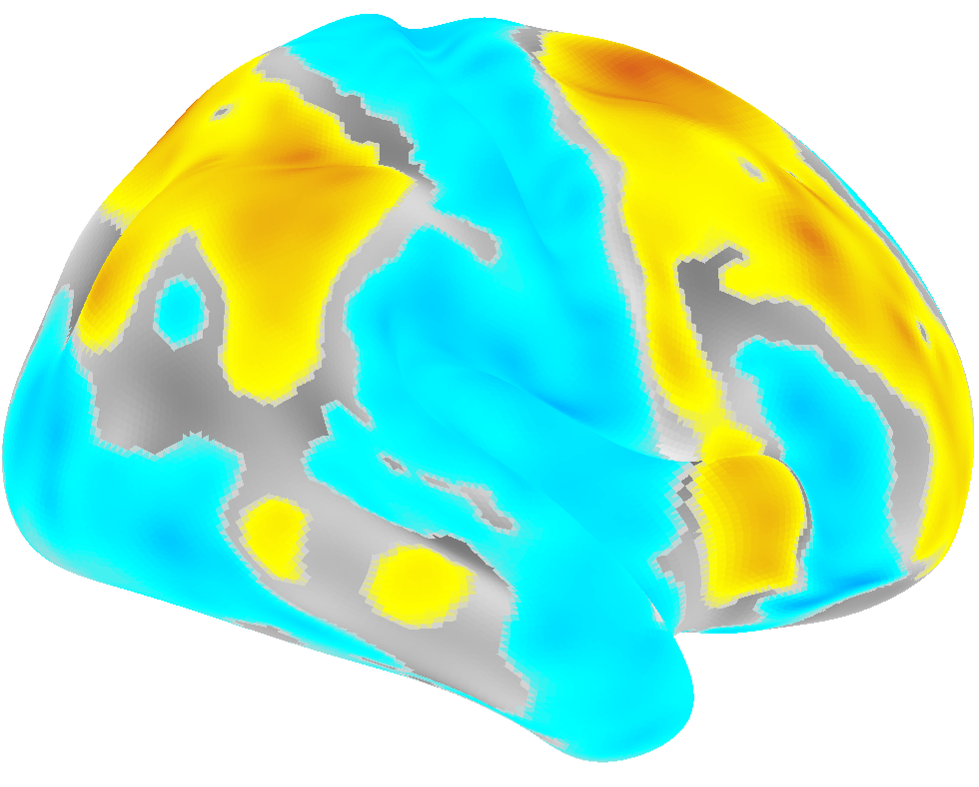}&\\
        \bottomrule
    \end{tabular}
     \hfill
    \caption{Estimated associations and selected activation regions between fMRI data and cognitive scores in the ABCD study for the left and right hemispheres. The values represent the expected change in the 2-back versus 0-back contrast (expressed as $z$-statistics) for a one-unit increase in cognitive score, while holding all other demographic information.}
    \label{fig:abcd_effect}
\end{figure}

In Figure~\ref{fig:abcd_effect}, we visualize the estimated associations between brain activity and cognitive function across both hemispheres using Nilearn \citep{Nilearn_contributors_Nilearn}. Results are again thresholded at $|E_{\rm s}|>0.95$. In this case, the extent of positive and negative regions are broadly similar across models (likely due to the large number of participants in this study), but detailed features differ. Compared with other models, SIMBA produces more spatially contiguous activation patterns with smooth boundaries and fewer very small islands, which makes them more interpretable. Notably, positive associations with the cognitive scores are observed in the dorsal attention network, fronto-parietal network, and salience network, which are the brain regions well known for supporting attentional control, cognitive flexibility, and integration of task-relevant information that are also align with established findings ~\citep{gordon2016generation, menon2011large}. Negative associations are identified in the cingulo-opercular, default mode, and parieto-occipital networks, consistent with suppression of task-irrelevant processes during working memory tasks.

\section{Discussion and Conclusion}\label{sec:discuss}
In this work, we propose SIMBA, a fully Bayesian spatial hierarchical model for whole brain analysis of task-based fMRI data. Our framework models population-level covariate effects and individual-specific deviations using Gaussian process priors, capturing spatially smooth and structured associations patterns across the brain. To mitigate the computational challenges of GP-based modeling for high-dimensional neuroimaging data, we utilize low-rank kernel approximation and a reparameterized representation that enables scalable posterior inference. We develop both a Gibbs sampling algorithm and an alternative based on variational inference, demonstrating relatively accurate posterior estimation and significant gains in computational efficiency. Comparing these two implementations of SIMBA, the variational inference method shows significantly reduced computational runtime, with similar and in several cases slightly improved model performance.

Through simulation studies, we show that the proposed SIMBA model outperforms classical voxel-wise GLM (i.e., a MUA) with standard FDR adjustment and BML models with information pooling in both estimation accuracy and signal detection power. This was quantitatively assessed using key relevant metrics, including both true positive and false discovery rates. Importantly, SIMBA typically had the leading counts for both simultaneously. SIMBA has the advantages of pooling spatial information at multiple scales, and, unlike standard MUA-based approaches, incorporates effect estimates (rather than being limited to solely adjusting statistics). Our model appears to more accurately reconstruct true spatial effects while also being more robust to the presence of noise.

Applications to real task-based fMRI datasets from the NARPS and ABCD studies further highlight the efficacy of SIMBA. Across both whole brain and cortical data, SIMBA identifies interpretable associations patterns that align with established findings, while also achieving improved model fitting that more align with true data distribution examined by PPC curves and predictive performance in out-of-sample evaluation metrics. These results demonstrate that SIMBA offers an accurate and computationally efficient approach for large-scale spatial modeling in neuroimaging, providing a powerful tool for detecting, quantifying, and interpreting brain activities across the whole brain volume.

One limitation of our proposed method is that the estimated effects are highly dependent on the choice of kernel functions. Although the Mat\'ern kernel is widely used for modeling associations functions, a parametric kernel function could still limit its flexibility. Exploring more flexible or nonparametric kernel representations could potentially improve the ability to capture complex, spatially varying functions in brain activity. Additionally, our use of low-rank kernel approximation is critical for computational scalability but inevitably introduces approximation error. While cross-validation helps in selecting an appropriate number of basis functions, an over-truncation may lead to underfitting, whereas including more bases increases computation demands or overfitting. In addition, for computational purposes, our model reparametrization requires the assumption that the main spatially varying effects share the same kernel structure with the spatial noise. This may limit the ability to capture different spatial dependence structures that could exist across these spatial components. Extending the model to incorporate more flexible or adaptive kernels would enhance its expressiveness but would also lead to additional computational costs. 

Another limitation arises in our current treatment of surface-based cortical gray matter fMRI data. Although the method is applied to data stored as meshes, we currently rely on geometric distances between vertices rather than explicitly modeling surface topology. This choice does not fully explore the rich structural information of surface-based data. Future work will aim to extend the framework to explicitly incorporate cortical surface topology into spatial dependencies, enabling more accurate modeling of brain activity on cortical manifolds.

There are several potential extensions of this work in both methodology and application that are worth exploring. Integrating SIMBA into the AFNI software suite would greatly increase its accessibility and practical utility for the neuroimaging community for a scalable Bayesian image regression pipeline. Work is currently underway for developing a command line tool for such distribution. Methodologically, one direction is to extend the SIMBA framework to incorporate within-individual or repeated-measures variables, enabling richer modeling of longitudinal task-based designs. Another is to account for measurement error in the response variable, for example by incorporating voxel-wise standard errors from participant-level analysis directly into the model. Finally, SIMBA could be adapted for voxel-level test–retest reliability estimation, providing a principled Bayesian approach to quantifying reproducibility in fMRI studies.

\section*{Data and Code Availability}
The NARPS datasets are publicly available online \url{https://www.narps.info/analysis.html}. We include the processed NARPS data and simulated data under an OSF directory stored at \url{https://osf.io/2d6t8/?view_only=933dc9343ebb4a13b9ee5b921bf46279}. The ABCD dataset can be accessed with permission from \url{https://abcdstudy.org}. The codebase for the SIMBA model and analysis scripts can be found at \url{https://github.com/y1zhong/apaper_simba}.
\section*{Author Contributions}
Y.Z: Methodology, software, formal analysis, visualization, writing—original draft preparation; G.C.: Methodology, conceptualization, writing—review and editing. P.A.T.: Methodology, visualization,  writing—review and editing; J.K: Conceptualization, methodology, supervision, writing—review and editing.

\section*{Funding}

JK was partially supported by NIH R01DA048993 and NIH R01MH105561.

JK and YZ was partially supported by NSF IIS2123777.

GC and PT were supported by the NIMH Intramural Research Programs (ZICMH002888) of the NIH (HHS, USA).

\section*{Declaration of Competing Interests}
The authors declare that they have no known competing financial interests in this paper.
\section*{Acknowledgements}

This research was supported in part by the Intramural Research Program of the National Institutes of Health (NIH). The contributions of the NIH authors were made as part of their official duties as NIH federal employees, are in compliance with agency policy requirements, and are considered Works of the United States Government. However, the findings and conclusions presented in this paper are those of the author(s) and do not necessarily reflect the views of the NIH or the U.S. Department of Health and Human Services.

\section*{Supplementary Material}\label{suppl_mat}
This supplementary document includes details of the posterior distributions of the full set of parameters: $\bfTheta=\{(\alpha_j)^J_{j=0}, (\bftheta_\beta)^J_{j=0}, (\boldsymbol{\theta}_{\eta_i})_{i=1}^N, \sigma_\alpha^2, \sigma_\beta^2, \sigma_\eta^2, \sigma_\epsilon^2, a_\alpha, a_\beta, a_\eta, a_\epsilon\}$, that are used to draw samples from Gibbs sampling algorithm in Section \ref{ss:gibbs}, and the variational distributions with the ELBO function in Section \ref{ss:VI}. The full code are implemented in Pytorch and included in the GitHub repository \url{https://github.com/y1zhong/apaper_simba}. 
\subsection*{Gibbs sampler}
Given the reparameterized model in \eqref{e:IMT} and the specifications of the conjugate priors, we present the posterior distributions for each parameter below.
Denote $\bfone_V\Phi = \widetilde{\Phi}$.\\
The full conditional distribution of each $\alpha_j, $ for $ j=0, \dots, J$ is
\begin{align*}
\alpha_j \mid \rest &\sim N\left (\mu_{\alpha_j}, \nu_{\alpha_j} \right );\\
\nu_{\alpha_j} &= \left( \frac{\sum_{i=1}^N x_{ij}^2\|\widetilde{\Phi}\|^2_F}{\sigma_\epsilon^2}  + \frac{1}{\sigma_\alpha^2} \right)^{-1}
\quad\\
\mu_{\alpha_j} &= \nu_{\alpha_j} \cdot \frac{1}{\sigma_\epsilon^2} \sum_{i=1}^N x_{ij} (\widetilde{\bfy}_i - \bfx_i^\top \bftheta_{\beta} - \bftheta_{\eta_i} \Phi_\eta)\widetilde{\Phi}^\top.
\end{align*}
The full conditional distribution of each $\bftheta_{\beta_j}$ for $ j=0, \dots, J$ is
\begin{align*}
\bftheta_{\beta_j} \mid \rest &\sim N\left (\mu_{\beta_j}, \Sigma_{\beta_j} \right );\\
\Sigma_{\beta_j} &= \left( \frac{\sum_{i=1}^N x_{ij}^2}{\sigma_\epsilon^2} + \frac{1}{\sigma_\beta^2} \right)^{-1}I_L\\
\mu_{\beta_j} &= \Sigma_{\beta_j} \cdot \frac{1}{\sigma_\epsilon^2} \sum_{i=1}^N x_{ij} \, (\widetilde{\bfy}_i - \bfx_i^\top\bfalpha \widetilde{\Phi} -\bftheta_{\eta_i} \Phi_\eta)
\end{align*}
The full conditional distribution of each $\bftheta_{\eta_i}$ for $ i=1, \dots, N$ is
\begin{align*}
\bftheta_{\eta_i} \mid \rest &\sim N\left (\mu_{\eta_i}, \Sigma_{\eta_i} \right );\\
\Sigma_{\eta_i} &=  \left(\frac{\Phi_\eta \Phi_\eta^\top}{\sigma^2_\epsilon}  +\frac{1}{\sigma^2_\eta}\right )^{-1}\\
\mu_{\eta_i} &= \Sigma_{\eta_i}\left(\frac{1}{\sigma^2_\epsilon}\left(\widetilde{\bfy}_i - \bfx_i^\top\bfalpha \widetilde{\Phi} -\bfx_i^\top \bftheta_{\beta}\right) \right)\Phi_\eta^\top\\
\end{align*}

The full conditional distribution of each variance parameter is
\begin{align*}
&\sigma_\epsilon^2 \mid \text{rest} \sim \mathrm{IG}\!\left( \frac{1 + NL}{2}, \; \frac{1}{2}\|\widetilde{\bfy}_i - \bfx_i^\top\bfalpha \widetilde{\Phi} - \bfx_i^\top \bftheta_{\beta} -\bftheta_{\eta_i} \Phi_\eta \|_2^2 + \frac{1}{a_\epsilon} \right), 
\quad
a_\epsilon \mid \text{rest} \sim \mathrm{IG}\!\left( 1, \; \frac{1}{A^2} + \frac{1}{\sigma_\epsilon^2} \right)\\
&\sigma_\eta^2 \mid \text{rest} \sim \mathrm{IG}\!\left( \frac{1 + N L_\eta}{2}, \; \frac{\sum^{N}_{i=1}\|\theta_{\eta_i}\|_2^2}{2} + \frac{1}{a_\eta} \right), 
\quad
a_\eta \mid \text{rest} \sim \mathrm{IG}\!\left( 1, \; \frac{1}{A^2} + \frac{1}{\sigma_\eta^2} \right)
\end{align*}
\begin{align*}
&\sigma_\beta^2 \mid \text{rest} \sim \mathrm{IG}\!\left( \frac{1 + (J+1) L}{2}, \; \frac{\sum^{J}_{j=0}\|\theta_{\beta_j}\|_2^2}{2} + \frac{1}{a_\beta} \right)\quad
a_\beta \mid \text{rest} \sim \mathrm{IG}\!\left( 1, \; \frac{1}{A^2} + \frac{1}{\sigma_\beta^2} \right)\\
&\sigma_\alpha^2 \mid \text{rest} \sim \mathrm{IG}\!\left( \frac{1 + J+1}{2}, \; \frac{\sum^{J}_{j=0}\alpha_j^2}{2} + \frac{1}{a_\alpha} \right), 
\quad
a_\alpha \mid \text{rest} \sim \mathrm{IG}\!\left( 1, \; \frac{1}{A^2} + \frac{1}{\sigma_\alpha^2} \right)
\end{align*}
\subsection*{Mathematical details in VI algorithm}
Under the framework of mean field variational inference, we approximate the posterior distributions using 
\begin{align*}
    p(\bfTheta|\bfX, \bfY)\approx q(\bfTheta)&= \prod^M_{m=1} q(z_m)\\
    &=\left( \prod_{j=0}^J q(\alpha_j) q(\bftheta_{\beta_j}) \right) \left(\prod_{i=1}^N q(\bftheta_{\eta_i}) \right) q(\sigma_\beta^2) q(\sigma_\eta^2) q(\sigma_\epsilon^2)q(a_\beta)q(a_\eta)q(a_\epsilon)
\end{align*}
For any $m$-th latent variable, the optimal $q(z_m)$ is given by:
\begin{equation*}
    q(z_m)\propto \text{exp}\left\{E_{-m}[\text{log}p(z_m|z_{-m}, \bfX, \bfY)]\right\}.
\end{equation*}
With derivation, we state the distributions under $q(\cdot)$ for each parameter as below:
\begin{align*}
\alpha_j \mid \rest &\sim N\left (\mu_{\alpha_j}, \nu_{\alpha_j} \right );\\
\nu_{\alpha_j} &= \left( \frac{\sum_{i=1}^N x_{ij}^2\|\widetilde{\Phi}\|^2_F}{E_q(\sigma_\epsilon^2)}  + \frac{1}{E_q(\sigma_\alpha^2)} \right)^{-1}
\quad\\
\mu_{\alpha_j} &= \nu_{\alpha_j} \cdot \frac{1}{E_q(\sigma_\epsilon^2)} \sum_{i=1}^N x_{ij} (\widetilde{\bfy}_i - \bfx_i^\top E_q(\bftheta_{\beta}) - E_q(\bftheta_{\eta_i} )\Phi_\eta)\widetilde{\Phi}^\top
\end{align*}

\begin{align*}
\bftheta_{\beta_j} \mid \rest &\sim N\left (\mu_{\beta_j}, \Sigma_{\beta_j} \right );\\
\Sigma_{\beta_j} &= \left( \frac{\sum_{i=1}^N x_{ij}^2}{E_q(\sigma_\epsilon^2)} + \frac{1}{E_q(\sigma_\beta^2)} \right)^{-1}I_L\\
\mu_{\beta_j} &= \Sigma_{\beta_j} \left( \frac{1}{E_q(\sigma_\epsilon^2)} \sum_{i=1}^N x_{ij} \, (\widetilde{\bfy}_i - \bfx_i^\top E_q(\bfalpha)\widetilde{\Phi} -E_q(\bftheta_{\eta_i}) \Phi_\eta)\right)
\end{align*}

\begin{align*}
\bftheta_{\eta_i} \mid \rest &\sim N\left (\mu_{\eta_i}, \Sigma_{\eta_i} \right );\\
\Sigma_{\eta_i} &=  \left(\frac{\Phi_\eta \Phi_\eta^\top}{E_q(\sigma^2_\epsilon)}  +\frac{1}{E_q(\sigma^2_\eta)}\right )^{-1}\\
\mu_{\eta_i} &= \Sigma_{\eta_i}\left(\frac{1}{E_q(\sigma^2_\epsilon)}\left(\widetilde{\bfy}_i - \bfx_i^\top E_q(\bfalpha) \widetilde{\Phi} -\bfx_i^\top E_q(\bftheta_{\beta})\right) \right)\Phi_\eta^\top\\
\end{align*}

\begin{align*}
\sigma_\epsilon^2 \mid \text{rest} &\sim \mathrm{IG}\!\left( \frac{1 + NL}{2}, \; \frac{1}{2}E_q \left(\|\widetilde{\bfy}_i - \bfx_i^\top\bfalpha \widetilde{\Phi} - \bfx_i^\top \bftheta_{\beta} -\bftheta_{\eta_i} \Phi_\eta \|_2^2\right) + \frac{1}{E_q(a_\epsilon)} \right) \\
a_\epsilon \mid \text{rest} &\sim \mathrm{IG}\!\left( 1, \; \frac{1}{A^2} + \frac{1}{E_q(\sigma_\epsilon^2)} \right)\\
\sigma_\eta^2 \mid \text{rest} &\sim \mathrm{IG}\!\left( \frac{1 + N L_\eta}{2}, \; \frac{\sum^{N}_{i=1}E_q(\|\theta_{\eta_i}\|_2^2)}{2} + \frac{1}{E_q(a_\eta)} \right) 
\\
a_\eta \mid \text{rest} &\sim \mathrm{IG}\!\left( 1, \; \frac{1}{A^2} + \frac{1}{E_q(\sigma_\eta^2)} \right)\\
\sigma_\beta^2 \mid \text{rest} &\sim \mathrm{IG}\!\left( \frac{1 + (J+1) L}{2}, \; \frac{\sum^{J}_{j=0}E_q(\|\theta_{\beta_j}\|_2^2)}{2} + \frac{1}{E_q(a_\beta)} \right)\\
a_\beta \mid \text{rest} &\sim \mathrm{IG}\!\left( 1, \; \frac{1}{A^2} + \frac{1}{E_q(\sigma_\beta^2)} \right)\\
\sigma_\alpha^2 \mid \text{rest} &\sim \mathrm{IG}\!\left( \frac{1 + J+1}{2}, \; \frac{\sum^{J}_{j=0}E_q(\alpha_j^2)}{2} + \frac{1}{E_q(a_\alpha)} \right) \\
a_\alpha \mid \text{rest} &\sim \mathrm{IG}\!\left( 1, \; \frac{1}{A^2} + \frac{1}{E_q(\sigma_\alpha^2)} \right)\\
\end{align*}
\printbibliography

@article{cox1996afni,
  title={{AFNI}: software for analysis and visualization of functional magnetic resonance neuroimages},
  author={Cox, Robert W},
  journal={Computers and Biomedical research},
  volume={29},
  number={3},
  pages={162--173},
  year={1996},
  publisher={Elsevier}
}

@article{gordon2016generation,
  title={Generation and evaluation of a cortical area parcellation from resting-state correlations},
  author={Gordon, Evan M and Laumann, Timothy O and Adeyemo, Babatunde and Huckins, Jeremy F and Kelley, William M and Petersen, Steven E},
  journal={Cerebral cortex},
  volume={26},
  number={1},
  pages={288--303},
  year={2016},
  publisher={Oxford University Press}
}

@article{gabry2019visualization,
  title={Visualization in {B}ayesian workflow},
  author={Gabry, Jonah and Simpson, Daniel and Vehtari, Aki and Betancourt, Michael and Gelman, Andrew},
  journal={Journal of the Royal Statistical Society Series A: Statistics in Society},
  volume={182},
  number={2},
  pages={389--402},
  year={2019},
  publisher={Oxford University Press}
}

@article{gelman1996posterior,
  title={Posterior predictive assessment of model fitness via realized discrepancies},
  author={Gelman, Andrew and Meng, Xiao-Li and Stern, Hal},
  journal={Statistica sinica},
  pages={733--760},
  year={1996},
  publisher={JSTOR}
}

@article{benjamini1995controlling,
  title={Controlling the false discovery rate: a practical and powerful approach to multiple testing},
  author={Benjamini, Yoav and Hochberg, Yosef},
  journal={Journal of the Royal statistical society: series B (Methodological)},
  volume={57},
  number={1},
  pages={289--300},
  year={1995},
  publisher={Wiley Online Library}
}

@inproceedings{Ansel_PyTorch_2_Faster_2024,
author = {Ansel, Jason and Yang, Edward and He, Horace and Gimelshein, Natalia and Jain, Animesh and Voznesensky, Michael and Bao, Bin and Bell, Peter and Berard, David and Burovski, Evgeni and Chauhan, Geeta and Chourdia, Anjali and Constable, Will and Desmaison, Alban and DeVito, Zachary and Ellison, Elias and Feng, Will and Gong, Jiong and Gschwind, Michael and Hirsh, Brian and Huang, Sherlock and Kalambarkar, Kshiteej and Kirsch, Laurent and Lazos, Michael and Lezcano, Mario and Liang, Yanbo and Liang, Jason and Lu, Yinghai and Luk, CK and Maher, Bert and Pan, Yunjie and Puhrsch, Christian and Reso, Matthias and Saroufim, Mark and Siraichi, Marcos Yukio and Suk, Helen and Suo, Michael and Tillet, Phil and Wang, Eikan and Wang, Xiaodong and Wen, William and Zhang, Shunting and Zhao, Xu and Zhou, Keren and Zou, Richard and Mathews, Ajit and Chanan, Gregory and Wu, Peng and Chintala, Soumith},
booktitle = {29th ACM International Conference on Architectural Support for Programming Languages and Operating Systems, Volume 2 (ASPLOS '24)},
doi = {10.1145/3620665.3640366},
month = apr,
publisher = {ACM},
title = {{PyTorch 2: Faster Machine Learning Through Dynamic {P}ython Bytecode Transformation and Graph Compilation}},
url = {https://docs.pytorch.org/assets/pytorch2-2.pdf},
year = {2024}
}

@article{casey2018adolescent,
  title={The {A}dolescent {B}rain {C}ognitive {D}evelopment ({ABCD}) study: {I}maging acquisition across 21 sites},
  author={Casey, Betty Jo and Cannonier, Tariq and Conley, May I and Cohen, Alexandra O and Barch, Deanna M and Heitzeg, Mary M and Soules, Mary E and Teslovich, Theresa and Dellarco, Danielle V and Garavan, Hugh and others},
  journal={Developmental Cognitive Neuroscience},
  volume={32},
  pages={43--54},
  year={2018},
  publisher={Elsevier}
}

@article{nystrom1930praktische,
  title={{\"U}ber die praktische Aufl{\"o}sung von Integralgleichungen mit Anwendungen auf Randwertaufgaben},
  author={Nystr{\"o}m, Evert J},
  year={1930}
}

@article{menon2011large,
  title={Large-scale brain networks and psychopathology: a unifying triple network model},
  author={Menon, Vinod},
  journal={Trends in cognitive sciences},
  volume={15},
  number={10},
  pages={483--506},
  year={2011},
  publisher={Elsevier}
}

@article{fischl2000measuring,
  title={Measuring the thickness of the human cerebral cortex from magnetic resonance images},
  author={Fischl, Bruce and Dale, Anders M},
  journal={Proceedings of the National Academy of Sciences},
  volume={97},
  number={20},
  pages={11050--11055},
  year={2000},
  publisher={The National Academy of Sciences}
}

@article{taylor2023highlight,
  title={Highlight results, don't hide them: Enhance interpretation, reduce biases and improve reproducibility},
  author={Taylor, Paul A and Reynolds, Richard C and Calhoun, Vince and Gonzalez-Castillo, Javier and Handwerker, Daniel A and Bandettini, Peter A and Mejia, Amanda F and Chen, Gang},
  journal={Neuroimage},
  volume={274},
  pages={120138},
  year={2023},
  publisher={Elsevier}
}

@article{ashburner2012spm,
  title={SPM: a history},
  author={Ashburner, John},
  journal={Neuroimage},
  volume={62},
  number={2},
  pages={791--800},
  year={2012},
  publisher={Elsevier}
}

@article{jenkinson2012fsl,
  title={Fsl},
  author={Jenkinson, Mark and Beckmann, Christian F and Behrens, Timothy EJ and Woolrich, Mark W and Smith, Stephen M},
  journal={Neuroimage},
  volume={62},
  number={2},
  pages={782--790},
  year={2012},
  publisher={Elsevier}
}

@article{jbabdi2009multiple,
  title={Multiple-subjects connectivity-based parcellation using hierarchical {D}irichlet process mixture models},
  author={Jbabdi, Saad and Woolrich, Mark William and Behrens, Timothy Edward John},
  journal={NeuroImage},
  volume={44},
  number={2},
  pages={373--384},
  year={2009},
  publisher={Elsevier}
}

@article{chen2019handling,
  title={Handling multiplicity in neuroimaging through {B}ayesian lenses with multilevel modeling},
  author={Chen, Gang and Xiao, Yaqiong and Taylor, Paul A and Rajendra, Justin K and Riggins, Tracy and Geng, Fengji and Redcay, Elizabeth and Cox, Robert W},
  journal={Neuroinformatics},
  volume={17},
  pages={515--545},
  year={2019},
  publisher={Springer}
}

@article{botvinik2020variability,
  title={Variability in the analysis of a single neuroimaging dataset by many teams},
  author={Botvinik-Nezer, Rotem and Holzmeister, Felix and Camerer, Colin F and Dreber, Anna and Huber, Juergen and Johannesson, Magnus and Kirchler, Michael and Iwanir, Roni and Mumford, Jeanette A and Adcock, R Alison and others},
  journal={Nature},
  volume={582},
  number={7810},
  pages={84--88},
  year={2020},
  publisher={Nature Publishing Group UK London}
}

@article{hartvig2000spatial,
  title={Spatial mixture modeling of f{MRI} data},
  author={Hartvig, Niels Vaever and Jensen, Jens Ledet},
  journal={Human brain mapping},
  volume={11},
  number={4},
  pages={233--248},
  year={2000},
  publisher={Wiley Online Library}
}

@article{woolrich2004constrained,
  title={Constrained linear basis sets for {HRF} modelling using Variational {B}ayes},
  author={Woolrich, Mark W and Behrens, Timothy EJ and Smith, Stephen M},
  journal={NeuroImage},
  volume={21},
  number={4},
  pages={1748--1761},
  year={2004},
  publisher={Elsevier}
}

@article{penny2005bayesian,
  title={{B}ayesian f{MRI} time series analysis with spatial priors},
  author={Penny, William D and Trujillo-Barreto, Nelson J and Friston, Karl J},
  journal={NeuroImage},
  volume={24},
  number={2},
  pages={350--362},
  year={2005},
  publisher={Elsevier}
}

@article{lin2024latent,
  title={Latent subgroup identification in image-on-scalar regression},
  author={Lin, Zikai and Si, Yajuan and Kang, Jian},
  journal={The annals of applied statistics},
  volume={18},
  number={1},
  pages={468},
  year={2024}
}

@article{whiteman2024bayesian,
  title={{B}ayesian inference for group-level cortical surface image-on-scalar regression with {G}aussian process priors},
  author={Whiteman, Andrew S and Johnson, Timothy D and Kang, Jian},
  journal={Biometrics},
  volume={80},
  number={4},
  pages={ujae116},
  year={2024},
  publisher={Oxford University Press}
}

@article{descombes1998spatio,
  title={Spatio-temporal f{MRI} analysis using {M}arkov random fields},
  author={Descombes, Xavier and Kruggel, Frithjof and Von Cramon, D Yves},
  journal={IEEE transactions on medical imaging},
  volume={17},
  number={6},
  pages={1028--1039},
  year={1998},
  publisher={IEEE}
}

@article{zhang2016spatiotemporal,
  title={A spatiotemporal nonparametric {B}ayesian model of multi-subject f{MRI} data},
  author={Zhang, Linlin and Guindani, Michele and Versace, Francesco and Engelmann, Jeffrey M and Vannucci, Marina},
  year={2016}
}

@article{siden2017fast,
  title={Fast {B}ayesian whole-brain f{MRI} analysis with spatial 3{D} priors},
  author={Sid{\'e}n, Per and Eklund, Anders and Bolin, David and Villani, Mattias},
  journal={NeuroImage},
  volume={146},
  pages={211--225},
  year={2017},
  publisher={Elsevier}
}

@article{siden2021spatial,
  title={Spatial 3{D} Mat{\'e}rn priors for fast whole-brain f{MRI} analysis},
  author={Sid{\'e}n, Per and Lindgren, Finn and Bolin, David and Eklund, Anders and Villani, Mattias},
  journal={Bayesian Analysis},
  volume={16},
  number={4},
  pages={1251--1278},
  year={2021},
  publisher={International Society for Bayesian Analysis}
}

@article{mejia2020bayesian,
  title={A {B}ayesian general linear modeling approach to cortical surface f{MRI} data analysis},
  author={Mejia, Amanda F and Yue, Yu and Bolin, David and Lindgren, Finn and Lindquist, Martin A},
  journal={Journal of the American Statistical Association},
  volume={115},
  number={530},
  pages={501--520},
  year={2020},
  publisher={Taylor \& Francis}
}

@article{salvatier2016probabilistic,
  title={Probabilistic programming in {P}ython using {P}y{MC}3},
  author={Salvatier, John and Wiecki, Thomas V and Fonnesbeck, Christopher},
  journal={PeerJ Computer Science},
  volume={2},
  pages={e55},
  year={2016},
  publisher={PeerJ Inc.}
}

@article{carpenter2017stan,
  title={Stan: A probabilistic programming language},
  author={Carpenter, Bob and Gelman, Andrew and Hoffman, Matthew D and Lee, Daniel and Goodrich, Ben and Betancourt, Michael and Brubaker, Marcus and Guo, Jiqiang and Li, Peter and Riddell, Allen},
  journal={Journal of statistical software},
  volume={76},
  pages={1--32},
  year={2017}
}

@article{gelman1992inference,
  title={Inference from iterative simulation using multiple sequences},
  author={Gelman, Andrew and Rubin, Donald B},
  journal={Statistical science},
  volume={7},
  number={4},
  pages={457--472},
  year={1992},
  publisher={Institute of Mathematical Statistics}
}

@article{marchini2004comparing,
  title={Comparing methods of analyzing f{MRI} statistical parametric maps},
  author={Marchini, Jonathan and Presanis, Anne},
  journal={Neuroimage},
  volume={22},
  number={3},
  pages={1203--1213},
  year={2004},
  publisher={Elsevier}
}

@article{reynolds_etal2024,
    author = {Reynolds, Richard C. and Glen, Daniel R. and Chen, Gang and Saad, Ziad S. and Cox, Robert W. and Taylor, Paul A.},
    title = {Processing, evaluating, and understanding F{MRI} data with afni\_proc.py},
    journal = {Imaging Neuroscience},
    volume = {2},
    pages = {imag-2-00347},
    year = {2024},
    month = {11},
    issn = {2837-6056},
    doi = {10.1162/imag_a_00347},
    url = {https://doi.org/10.1162/imag\_a\_00347},
    eprint = {https://direct.mit.edu/imag/article-pdf/doi/10.1162/imag\_a\_00347/2479365/imag\_a\_00347.pdf},
}

@article{taylor_etal2024,
    author = {Taylor, Paul A. and Glen, Daniel R. and Chen, Gang and Cox, Robert W. and Hanayik, Taylor and Rorden, Chris and Nielson, Dylan M. and Rajendra, Justin K. and Reynolds, Richard C.},
    title = {A Set of F{MRI} Quality Control Tools in AFNI: Systematic, in-depth, and interactive {QC} with afni\_proc.py and more},
    journal = {Imaging Neuroscience},
    volume = {2},
    pages = {imag-2-00246},
    year = {2024},
    month = {08},
    issn = {2837-6056},
    doi = {10.1162/imag_a_00246},
    url = {https://doi.org/10.1162/imag\_a\_00246},
    eprint = {https://direct.mit.edu/imag/article-pdf/doi/10.1162/imag\_a\_00246/2464165/imag\_a\_00246.pdf},
}

@article{allen_etal2012,
title = {Data Visualization in the Neurosciences: Overcoming the Curse of Dimensionality},
journal = {Neuron},
volume = {74},
number = {4},
pages = {603-608},
year = {2012},
issn = {0896-6273},
doi = {https://doi.org/10.1016/j.neuron.2012.05.001},
url = {https://www.sciencedirect.com/science/article/pii/S089662731200428X},
author = {Elena A. Allen and Erik B. Erhardt and Vince D. Calhoun},
}

@ARTICLE{sripada2020prediction,
  title     = "Prediction of neurocognition in youth from resting state {fMRI}",
  author    = "Sripada, Chandra and Rutherford, Saige and Angstadt, Mike and
               Thompson, Wesley K and Luciana, Monica and Weigard, Alexander
               and Hyde, Luke H and Heitzeg, Mary",
  journal   = "Mol. Psychiatry",
  publisher = "Springer Science and Business Media LLC",
  volume    =  25,
  number    =  12,
  pages     = "3413--3421",
  month     =  dec,
  year      =  2020,
  language  = "en"
}

@ARTICLE{Chen2017-qt,
  title     = "Is the statistic value all we should care about in neuroimaging?",
  author    = "Chen, Gang and Taylor, Paul A and Cox, Robert W",
  journal   = "Neuroimage",
  publisher = "Elsevier BV",
  volume    =  147,
  pages     = "952--959",
  month     =  feb,
  year      =  2017,
  language  = "en"
}

@software{Nilearn_contributors_Nilearn,
author = "{Nilearn contributors}",
title = {{Nilearn}},
year =  2025,
url = {https://doi.org/10.5281/zenodo.8397156}
}




\end{document}